\newif\ifTechRep
\TechReptrue

\newcommand\asd{\textsc{AID}\xspace}
\newcommand\asdFullAllCap{Adaptive Interventional Debugging\xspace}

\documentclass{vldb}
\usepackage{graphicx}
\usepackage{balance} 
\usepackage{enumitem}
\usepackage{times}

\usepackage{amsthm, bm}
\usepackage[ruled,vlined]{algorithm2e}
\usepackage[labelfont=bf]{caption}
\usepackage[labelformat=simple]{subcaption}
\usepackage{xspace}
\usepackage{booktabs}
\usepackage{url}
\usepackage{multicol}
\usepackage{multirow}
\usepackage{color}
\usepackage{hyperref}
\usepackage[usernames,dvipsnames,svgnames,table]{xcolor}
\usepackage[utf8]{inputenc}
\usepackage{gensymb}

\newcommand{\stepCounter}[1]{\raisebox{.5pt}{\textcircled{\raisebox{-.9pt}{#1}}}}
\definecolor{vlightgray}{gray}{0.85}
\setlist{nolistsep,leftmargin=*}

\newtheorem{example}{Example}
\newtheorem{example-summary}{Example Summary}

\newtheorem{definition}{Definition}
\newtheorem{theorem}{Theorem}
\newtheorem{lemma}[theorem]{Lemma}

\let\paragraph\relax 
\newcommand{\paragraph}[1]{\noindent\textbf{#1}}

\renewcommand{\footnotesize}{\scriptsize}

\hypersetup{%
  bookmarksdepth = {-2},
  pdftitle = {Causality-Guided Adaptive Interventional Debugging},
  pdfkeywords = {},
  pdfauthor = {},
  pdfstartview={FitH},
  urlcolor=black,
  linkcolor=black,
  citecolor=black,
  colorlinks=true,
}

\makeatletter
\def\@copyrightspace{\relax}
\makeatother

\makeatletter
\def\@mkbibcitation{\relax}
\makeatother

\vldbTitle{AID}
\vldbAuthors{}
\vldbDOI{https://doi.org/10.14778/xxxxxxx.xxxxxxx}
\vldbVolume{12}
\vldbNumber{xxx}
\vldbYear{2020}

\SetCommentSty{mycommfont}
\setlist{nolistsep,leftmargin=*}
\theoremstyle{plain}

\renewcommand{\footnotesize}{\scriptsize}

\definecolor{teal}{RGB}{0,150,0}

\newcommand{\teal}{black}
\newcommand{\blue}{black}
\newcommand{\red}{black}

\newcommand{\revisefive}[1]{{\color{\teal} #1}}
\newcommand{\revisesix}[1]{{\color{\blue} #1}}
\newcommand{\reviseseven}[1]{{\color{\red} #1}}

\newcommand\acm{AC-DAG\xspace}
\newcommand\acmFull{approximate causal DAG\xspace}
\newcommand\acmChain{approximate causal chain\xspace}
\newcommand\acmFullAllCap{Approximate Causal DAG\xspace}
\newcommand\acmFullFirstCap{Approximate causal DAG\xspace}

\newcommand{\causalPath}{\mathbb{P}}
\newcommand{\Vt}{\mathcal{V}}
\newcommand{\Et}{\mathcal{E}}
\newcommand{\DAG}{\mathcal{G}}
\newcommand{\Preds}{\mathcal{P}}
\newcommand{\cSet}{\mathcal{C}}
\newcommand{\xSet}{\mathcal{X}}
\newcommand{\branch}{\mathcal{B}}

\newcommand{\techRepCitation}{\ifTechRep  (see Appendix)\xspace\else \cite{technicalReport}\xspace\fi}
\newcommand{\citeTechRep}{\cite{technicalReport}}

\newcommand{\appOrTechRep}{\ifTechRep the Appendix\xspace\else our technical report~\citeTechRep\xspace\fi}

\SetKw{Break}{break}
\SetKwInOut{Input}{Input}
\SetKwInOut{Output}{Output}
\SetKwRepeat{Do}{do}{while}
\SetKw{Break}{break}

\begin{document}
	
\title{\mbox{Causality-Guided Adaptive Interventional Debugging}}
\subtitle{Technical Report}

\numberofauthors{3} 
\author{
\alignauthor
Anna Fariha\\
       \affaddr{University of Massachusetts}
	   \affaddr{Amherst, MA}\\
       \email{\large afariha@cs.umass.edu}
\alignauthor
Suman Nath\\
	\affaddr{Microsoft Research}\\
	\affaddr{Redmond, WA}\\
	\email{\large Suman.Nath@microsoft.com}
\alignauthor
 Alexandra Meliou\\
       \affaddr{University of Massachusetts}
	   \affaddr{Amherst, MA}\\
       \email{\large ameli@cs.umass.edu}
}

\maketitle

\begin{abstract}
Runtime nondeterminism is a fact of life in modern database applications.
Previous research has shown that nondeterminism can cause applications to {\em
intermittently} crash, become unresponsive, or experience data corruption. We
propose \asdFullAllCap (\asd) for debugging such intermittent
failures.

\begin{sloppypar} \asd combines existing statistical debugging, causal
analysis, fault injection, and group testing techniques in a novel way to
(1)~pinpoint the root cause of an application's intermittent failure and
(2)~generate an explanation of how the root cause triggers the failure. \asd
works by first identifying a set of runtime behaviors (called predicates) that
are strongly correlated to the failure. It then utilizes temporal properties of
the predicates to (over)-approximate their causal relationships. Finally, it
uses fault injection to execute a sequence of interventions on the predicates
and discover their true causal relationships. This enables \asd to identify the
true root cause and its causal relationship to the failure. We theoretically
analyze how fast \asd can converge to the identification.

We evaluate \asd with six real-world applications that intermittently fail
under specific inputs. In each case, \asd was able to identify the root cause
and explain how the root cause triggered the failure, much faster than group
testing and more precisely than statistical debugging. We also evaluate \asd
with many synthetically generated applications with known root causes and
confirm that the benefits also hold for them. \end{sloppypar}

\end{abstract}

\keywords{Root cause analysis; trace analysis; causal intervention; group
testing; concurrency bug}

\hypersetup{%
  bookmarksdepth = {-2},
  pdftitle = {Causality-Guided Adaptive Interventional Debugging},
  pdfkeywords = {},
  pdfauthor = {},
  pdfstartview={FitH},
  urlcolor=black,
  linkcolor=black,
  citecolor=black,
  colorlinks=true,
}

\section{Introduction}\label{sec:intro}
Modern data management systems and database-backed applications
run on commodity hardware and heavily rely on asynchronous and concurrent
processing~\cite{DBLP:journals/cacm/DeanG08,DBLP:journals/sigops/Herlihy92,
DBLP:conf/mss/ShvachkoKRC10, mysql}. As a result, they commonly experience 
runtime nondeterminism such as transient faults and variability in timing and 
thread scheduling. Unfortunately, software bugs related to handling 
nondeterminism are also common to these systems. Previous studies reported 
such bugs in MySQL~\cite{DBLP:conf/asplos/LuPSZ08, bovenzi2012aging},
PostgreSQL~\cite{lu2007muvi}, NoSQL systems~\cite{leesatapornwongsa2016taxdc,
yuan2014simple}, and database-backed applications~\cite{bailis2015feral}, and
showed that the bugs can cause crashes, unresponsiveness, and data corruptions.
It is, therefore, crucial to identify and fix these bugs as early as possible.

Unfortunately, localizing root causes of intermittent failures is extremely
challenging~\cite{luo2014empirical, DBLP:conf/icml/ZhengJLNA06,
DBLP:conf/icsm/LiuQWM14}. For example, concurrency bugs such as deadlocks,
order and atomicity violation, race conditions, etc. may appear only under very
specific thread interleavings. Even when an application executes with the same
input in the same environment, these bugs may appear only rarely (e.g., in {\em
flaky} unit tests~\cite{luo2014empirical}). When a. When a concurrency bug is
confirmed to exist, the debugging process is further complicated by the fact
that the bug cannot be consistently reproduced. Heavy-weight techniques based
on record-replay~\cite{DBLP:conf/osdi/AttariyanCF12} and fine-grained tracing
with lineage~\cite{DBLP:conf/sigmod/AlvaroRH15, DBLP:conf/cidr/OldenburgZRA19}
can provide insights on root causes after a bug manifests; but their runtime
overheads often interfere with thread timing and scheduling, making it even
harder for the intermittent bugs to manifest in the first
place~\cite{lam2019root}.

{\em Statistical Debugging} (SD)~\cite{DBLP:conf/kbse/JonesH05,
statisticalDebuggingLiblit, Sober, crugLiblit} is a data-driven technique that
partly addresses the above challenge. SD uses lightweight logging to capture an
application's runtime (mis)\-behaviors, called {\em predicates}. An example
predicate indicates whether a method returns null in a particular execution or
not. Given an application that intermittently fails, SD logs predicates from
many successful and failed executions. SD then uses statistical analyses of the
logs to identify {\em discriminative predicates} that are highly correlated
with the failure.

\looseness-1 SD has two key limitations. First, SD can produce many
discriminative predicates that are correlated to, but not a true cause of, a
failure. Second, SD does not provide enough insights that can explain how a
predicate may eventually lead to the failure. Lack of such insights and the
presence of many non-causal predicates make it hard for a developer to identify
the true root cause of a failure. SD expects that a developer has sufficient
domain knowledge about if/how a predicate can eventually cause a failure, even
when the predicate is examined in isolation without additional context. This is
often hard in practice, as is reported by real-world
surveys~\cite{DBLP:conf/issta/ParninO11}.
\begin{example}\label{ex:npgsql}
	\sloppy
	 To motivate our work, we consider a recently reported issue in
	 Npgsql~\cite{npgsql}, an open-source ADO.NET data provider for PostgreSQL. On
	 its GitHub repository, a user reported that a database application
	 intermittently crashes when it tries to create a new PostgreSQL connection
	 (GitHub issue \#2485~\cite{npgsqlBug2485}). The underlying root cause is a
	 data race on an array index variable. The data race, which happens only when
	 racing threads interleave in a specific way, causes one of the threads to
	 access beyond the size of the array. This causes an exception that crashes
	 the application.

	 We used SD to localize the root cause of this nondeterministic bug (more
	 details are in Section~\ref{sec:experimental-evaluation}). SD identified 14
	 predicates, only three of which were causally related to the error. Other
	 predicates were just symptoms of the root cause or happened to co-occur with
	 the root cause.
\end{example}

\looseness-1 In Section~\ref{sec:experimental-evaluation}, we describe five
other case studies that show the same general problem: SD produces too many
predicates, only a small subset of which are causally related to the failure.
Thus, SD is not specific enough, and it leaves the developer with the task of
identifying the root causes from a large number of candidates. This task is
particularly challenging, since SD does not provide explanations of how a
potential predicate can eventually lead to the failure.

\looseness-1 In this paper, we address these limitations with a new data-driven 
technique called {\em \asdFullAllCap} (\asd). Given predicate logs from 
successful and failed
executions of an application, \asd can pinpoint {\em why} the application
failed, by identifying one (or a small number of) predicate that indicates the
real root cause (instead of producing a large number of potentially unrelated
predicates). Moreover, \asd can explain {\em how} the root cause leads to the
failure, by automatically generating a causal chain of predicates linking the
root cause, subsequent effects, and the failure. By doing so, \asd enables a
developer to quickly localize (and fix) the bug, even without deep knowledge
about the application.

\looseness-1 \asd achieves the above by combining SD with causal
analysis~\cite{DBLP:journals/pvldb/MeliouRS14, DBLP:conf/sigmod/MeliouGNS11,
DBLP:conf/mud/MeliouGMS10}, fault injection~\cite{DBLP:conf/sigmod/AlvaroRH15,
han1995doctor, DBLP:journals/tc/KanawatiKA95}, and group
testing~\cite{hwang-group-testing} in a novel way. Like SD, it starts by
identifying discriminative predicates from successful and failed executions. In
addition, \asd uses temporal properties of the predicates to build an {\em
\acmFull} (Directed Acyclic Graph), which contains a superset of all true
causal relationships among predicates. \asd then progressively refines the DAG.
In each round of refinement, \asd uses ideas from adaptive group testing to
carefully select a subset of predicates. Then, \asd re-executes the application
during which it \emph{intervenes} (i.e., modifies application's behavior by
e.g., injecting faults) application to fail or not, \asd confirms or discards
causal relationships in the \acmFull, assuming counterfactual causality ($C$ is
a counterfactual cause of $F$ iff $F$ would not occur unless $C$ occurs) and a
single root cause. A sequence of interventions enables \asd to identify the
root cause and generate a \emph{causal explanation path}, a sequence of
causally-related predicates that connect the root cause to the failure.

A key benefit of \asd is its efficiency---it can identify root-cause and
explanation predicates with significantly fewer rounds of interventions than
adaptive group testing. In group testing, predicates are considered independent
and hence each round selects {\em a random subset} of predicates to
intervene on and makes causality decisions about only those intervened
predicates. In contrast, \asd uses potential causality among predicates (in the
\acmFull). This enables \asd to (1)~make decisions not only about the intervened
predicates, but also about other predicates; and (2)~carefully select
predicates whose intervention would maximize the effect of (1).
Through theoretical and empirical analyses we show that this can significantly
reduce the number of required interventions. This is an important benefit in
practice since each round of intervention involves executing the application
with fault injection and hence is time-consuming.

\sloppy
We evaluated \asd on 3 open-source applications: Npgsql, Apache Kafka, 
Microsoft Azure Cosmos DB, and on 3 proprietary applications in Microsoft. 
We used known issues that cause these applications to intermittently fail even 
for same inputs. In each case, \asd was able to identify the root cause of 
failure and generate an explanation that is consistent with the explanation 
provided by respective developers. Moreover, \asd achieved this with 
significantly fewer interventions than  traditional adaptive group testing. We  
also performed sensitivity analysis of \asd with a set of synthetic workloads. 
The results show that \asd requires fewer interventions than traditional 
adaptive group testing, and has significantly better worst-case performance 
than other variants.

In summary, we make the following contributions: 
\begin{itemize}
	 \item We propose \asdFullAllCap (\asd), a data-driven technique that
	 localizes the root cause of an intermittent failure through a novel combination
	 of statistical debugging, causal analysis, fault injection, and group testing
	 (Section~\ref{sec:background}). \asd provides significant benefits over the
	 state-of-the-art Statistical Debugging (SD) techniques by (1)~pinpointing the
	 root cause of an application's failure and (2)~generating an explanation of
	 how the root cause triggers the failure
	 (Sections~\ref{sec:asd}--\ref{sec:refining-causality}). In contrast, SD
	 techniques generate a large number of potential causes and without
	 explaining how a potential cause may trigger the failure.
	
	 \item We use information theoretic analysis to show that \asd, by utilizing
	 causal relationship among predicates, can converge to the true root cause and
	 explanation significantly faster than traditional adaptive group testing
	 (Section~\ref{sec:complexity-analysis}).
	
	 \item We evaluate \asd with six real-world applications that
	 intermittently fail under specific inputs
	 (Section~\ref{sec:experimental-evaluation}). \asd was able to identify the
	 root causes and explain how the root causes triggered the failure, much
	 faster than adaptive group testing and more precisely than SD. We also
	 evaluate \asd with many synthetically generated applications with known
	 root causes and confirm that the benefits hold for them as well.

\end{itemize}

\section{Background and Preliminaries}\label{sec:background}

\asd combines several existing techniques in a novel way. We now briefly review
the techniques.

\subsubsection*{Statistical Debugging} 
Statistical debugging (SD) aims to automatically pinpoint likely causes for an
application's failure by statistically analyzing its execution logs from many
successful and failed executions. It works by instrumenting an application to
capture runtime {\em predicates} about the application's behavior. Examples of
predicates include ``the program takes the \texttt{false} branch at line 31'',
``the method \texttt{foo()} returns \texttt{null}'', etc. Executing the
instrumented application generates a sequence of predicate values, which we
refer to as {\em predicate logs}. Without loss of generality, we assume that
all predicates are Boolean.

\looseness-1 Intuitively, the true root cause of the failure will cause certain
predicates to be true only in the failed logs (or, only in the successful
logs). Given logs from many successful executions and many failed executions of
an application, SD aims to identify those {\em discriminative} predicates.
Discriminative predicates encode program behaviors of failed executions that
deviate from the ideal behaviors of the successful executions. Without loss of
generality, we assume that discriminative predicates are {\tt true} during
failed executions. The predicates can further be ranked based on their
\emph{precision} and \emph{recall}, two well-known metrics that capture their
discriminatory power.
\begin{align*}
precision(P) &= \frac{\text{\#failed executions where $P$  is  \texttt{true}}}{\text{\#executions where $P$ is \texttt{true}}}\\
recall(P) &= \frac{\text{\#failed executions where $P$ is \texttt{true}}}{\text{\#failed executions}}
\end{align*}

\subsubsection*{Causality}
Informally, causality characterizes the relationship between an event and an
outcome: the event is a cause if the outcome is a consequence of the event.
There are several definitions of causality~\cite{HP01,
DBLP:journals/amai/Pearl11}. In this work, we focus on \emph{counterfactual}
causes. According to \emph{counterfactual causality}, C causes E iff E would
not occur unless C occurs.
Reasoning about causality frequently relies on a mechanism for
\emph{interventions}~\cite{Pearl2000, Hitchcock2015, woodward2003making,
Spirtes2000}, where one or more variables are forced to particular values,
while the mechanisms controlling other variables remain unperturbed. Such
interventions uncover counterfactual dependencies between variables.

\looseness-1 Trivially, executing a program is a cause of its failure: if the
program was not executed at the first place, the failure would not have
occurred. However, our analysis targets \emph{fully}-discriminative predicates
(with 100\% precision and 100\% recall), thereby eliminating such trivial
predicates that are program invariants.

\subsubsection*{Fault Injection}
In software testing, {\em fault
injection}~\cite{DBLP:conf/sigmod/AlvaroRH15, han1995doctor,
DBLP:journals/tc/KanawatiKA95, marinescu2009lfi} is a technique to force an
application, by instrumenting it or by manipulating the runtime environment, to
execute a different code path than usual. We use the technique to {\em
intervene} on (i.e., repair) discriminative predicates. Consider a method
\texttt{ExecQuery()} that returns a result object in all successful executions
and \texttt{null} in all failed executions. Then, the predicate
``\texttt{ExecQuery()} returns \texttt{null}'' is discriminative. The predicate
can be intervened by forcing \texttt{ExecQuery()} to return the correct result
object. Similarly, the predicate ``there is a data race on X'' can be
intervened by delaying one access to X or by putting a lock around the code
segments that access X to avoid simultaneous accesses to X.

\subsubsection*{Group Testing}
Given a set of discriminative predicates, a na\"ive approach to identify which
predicates cause the failure is to intervene on one predicate at a time and
observe if the intervention causes an execution to succeed. However, the
number of required interventions is linear in number of predicates. {\em Group
testing} reduces the number of interventions.

\looseness-1 Group testing refers to the procedure that identifies certain
items (e.g., defective) among a set of items while minimizing the number of
\emph{group tests} required. Formally, given a set $\Preds$ of $N$ elements
where $D$ of them are defective, group testing performs $k$ group tests, each
on group $\Preds_i \subseteq \Preds$. Result of test on group $\Preds_i$ is
positive if $\exists P \in \Preds_i \text{ s.t. $P$ is defective}$, and
negative otherwise. The objective is to minimize $k$, i.e., the number of group
tests required. In our context, a group test is simultaneous intervention on a
group of predicates, and the goal is to identify the predicates that cause the
failure.

Two variations of group testing are studied in the literature: \emph{adaptive}
and \emph{non-adaptive}. Our approach is based on adaptive group testing where
the $i$-th group to test is decided {\em after} we observe the results of all 
$1 \le j < i$ previous group tests. 
A trivial upper bound for adaptive group testing~\cite{hwang-group-testing} is 
$\mathcal{O}(D\log N)$. A simple binary search algorithm can find each of the 
$D$ defective items in at most $\log N$ group tests and hence a total of $D 
\log N$ group tests are sufficient to identify all defective items. Note that 
if $D \ge \frac{N}{\log N}$, then a linear strategy is preferable over any 
group testing scheme. Hence, we assume that $D < \frac{N}{\log N}$.

\section{AID Overview}\label{sec:asd} 

\asdFullAllCap(\asd) targets applications (e.g., {\em flaky} 
tests~\cite{luo2014empirical}) that, even 
with the same inputs, intermittently fail due to various runtime nondeterminism 
such as thread scheduling and timing. Given predicate logs of successful and 
failed executions of an application, the goals of \asd are to (1)~identify {\em 
what} predicate actually causes the failure, and (2)~generate an explanation of 
{\em how} the root cause leads to the failure (via a sequence of intermediate 
predicates). This is in contrast with traditional statistical debugging, which 
generates a set of potential root-cause predicates (often a large number),
without any explanation of how each potential root cause may lead to the 
failure.

\begin{figure}[t]
	\begin{center}
		\includegraphics[width=0.47\textwidth]{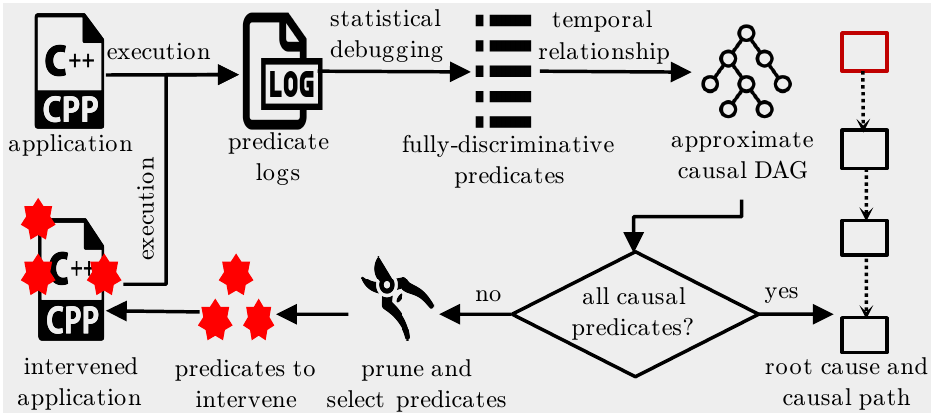}
		\caption{\asdFullAllCap workflow.} 
		\vspace{-5mm}
		\label{architecture}
	\end{center}
\end{figure}

Figure~\ref{architecture} shows an overview of \asd. First, the framework
employs standard SD techniques on predicate logs to identify a set of
\emph{fully}-discriminative predicates, i.e., predicates that \emph{always}
appear in the failed executions and \emph{never} appear in the successful
executions. Then, \asd uses the temporal relationships of predicates to infer 
\emph{approximate causality}: if $P_1$ temporally precedes $P_2$ in 
\emph{all} logs where they both appear, then $P_1$ \emph{may} cause $P_2$. \asd 
represents this approximate causality in a DAG called {\em \acmFullAllCap} (\acm), 
where predicates are nodes and edges indicate these possible causal 
relationships. We describe the \acm in Section~\ref{sec:ac}.

\looseness-1
Based on its construction, the \acm is guaranteed to contain all the true
root-cause predicates and causal relationships among predicates. However, it
may also contain additional predicates and edges that are not truly causal. The
key insight of \asd is that we can refine the \acm and prune the non-causal
nodes and edges through a sequence of interventions. \reviseseven{To intervene
on a predicate, \asd changes the application's behavior through fault injection so that the
predicate's value matches its value in successful executions. If the failure
does not occur under the intervention, then, based on counterfactual
causality, the predicate is
guaranteed to be a root cause of the failure.} Over several iterations, \asd
intervenes on a set of carefully chosen predicates, refines the set of
discriminative predicates, and prunes the \acm, until it discovers the true
root cause and the path that leads to the failure. We describe the intervention
mechanism of \asd in Section~\ref{sec:refining-causality}.

\smallskip 

We now describe how \asd adapts existing approaches in SD and fault injection for
two of its core ideas: predicates and interventions. We refer to
\appOrTechRep for additional details and discussion.

\revisefive{ 

\subsection{AID Predicates}\label{sec:asd-predeicates}
 
\noindent 
\textbf{Predicate design:} 
Similar to traditional SD
techniques, \asd is effective only if the initial set of predicates (in
the predicate logs) contains a {\em root-cause predicate} that causes the
failure. Predicate design is orthogonal to \asd. We use predicates
used by existing SD techniques, especially the ones used for finding
root causes of concurrency bugs~\cite{crugLiblit}, a key reason behind
intermittent failures~\cite{luo2014empirical}.
Figure~\ref{tab:predicatesAndFIs} shows examples of predicates in \asd (column
1).

\smallskip 
\noindent 
\textbf{Predicate extraction:}
\asd automatically instruments a target application to generate its
{\em execution trace}~\techRepCitation.
The trace contains each executed method's
start and end time, its thread id, ids of objects it accesses, return values,
whether it throws exception or not, and so on. This trace is then analyzed
offline to evaluate a set of predicates at each execution point. This results
in a sequence of predicates, called {\em predicate log}. The instrumented
application is executed multiple times with the same input, to generate a set
of predicate logs, each labeled as a successful or failed execution.
Figure~\ref{tab:predicatesAndFIs} shows the runtime conditions used
to extract predicates (column 2).

\smallskip 
\noindent 
\textbf{Modeling nondeterminism:} 
In practice, some predicates may cause a failure nondeterministically: two
predicates A and B in \emph{conjunction} cause a failure. \asd does not
consider such predicates since they are not fully discriminative (recall $<$
100\%). However, \asd can still model these cases with \emph{compound}
predicates, adapted from state-of-the-art SD techniques~\cite{crugLiblit},
which model conjunctions. These compound predicates (``A and B'') would
\emph{deterministically} cause the failure and hence be fully discriminative.
Note that \asd focuses on counterfactual causality and thus does not support
\emph{disjunctive} root causes (as they are not counterfactual).
In Section~\ref{sec:refining-causality}, we discuss \asd's assumptions and
their impact in practice.
}

\begin{figure*}[t]
	\renewcommand{\arraystretch}{1.2}
	{
		\setlength\tabcolsep{3pt}
		\begin{tabular}{|p{4cm}|p{6.8cm}|p{6cm}|}		
			\hline
			\multicolumn{1}{|c}{\color{\teal}\textbf{(1) Predicate}} &
			\multicolumn{1}{|c}{\color{\teal}\textbf{(2) Extraction condition}} & 
			\multicolumn{1}{|c|}{\color{\red}\textbf{(3) Intervention mechanism}}\\
			\hline
			\hline
			\textcolor{\teal}{There is a data race involving methods $M_1$ and $M_2$}& 
			\textcolor{\teal}{$M_1$ and $M_2$ temporally overlap accessing some object $X$ while one of them is a \texttt{write}} &
			\textcolor{\red}{Put locks around the code segments within $M_1$ and $M_2$ that access $X$}\\
			\hline
			\textcolor{\teal}{Method $M$ fails} &
			\textcolor{\teal}{$M$ throws an exception} &
			\textcolor{\red}{Put $M$ in a \texttt{try-catch} block}\\
			\hline
			\textcolor{\teal}{Method $M$ runs too fast} & 
			\textcolor{\teal}{$M$'s duration is less than the minimum duration for $M$ among all successful executions} &
			\textcolor{\red}{Insert delay before $M$'s \texttt{return} statement}\\
			\hline
			\textcolor{\teal}{Method $M$ runs too slow}& 
			\textcolor{\teal}{$M$'s duration is greater than the maximum duration for $M$ among all successful executions} &
			\textcolor{\red}{Prematurely return from $M$ the correct value that $M$ returns in all successful executions}\\
			\hline
			\textcolor{\teal}{Method $M$ returns incorrect value}&
			\textcolor{\teal}{$M$'s return value $ \neq x$, where $x$ is the correct value returned by $M$ in all successful executions}&
			\textcolor{\red}{Alter $M$'s \texttt{return} statement to force it to return the correct value $x$}\\		
			\hline
		\end{tabular}
	}
	\vspace{-1mm}
	\caption{\revisefive{Few example predicates, conditions used to 
			extract them, 
			and the \reviseseven{corresponding interventions using fault injection.}}}
	\label{tab:predicatesAndFIs}
\end{figure*}

\reviseseven{ 

\subsection{AID Interventions}\label{sec:aid-intervention}

\noindent
\textbf{Intervention mechanism:} 
\asd uses an existing fault injection tool (similar to
LFI~\cite{marinescu2009lfi}) to intervene on fully-discrimina\-tive predicates;
interventions change a predicate to match its value in a successful execution.
In
a way, \asd's interventions try to locally ``repair'' a failed execution.
Figure~\ref{tab:predicatesAndFIs} shows examples of \asd's interventions (column 3). Most of the interventions rely on changing timing and
thread scheduling that can occur naturally by the underlying execution environment
and runtime. More specifically, \asd can slow down the execution of a method (by
injecting delays), force or prevent concurrent execution of methods in
different threads (by using synchronization primitives such as locks), change
the execution order of concurrent threads (by injecting delays), etc. Such
interventions can repair many concurrency bugs.
}

\reviseseven{

\smallskip 
\noindent 
\begin{sloppypar}
\textbf{Validity of intervention:} 
\asd supports two additional intervention types, return-value alteration and
exception-handling, which, in theory, can have undesirable runtime
side-effects. Consider two predicates: (1)~method
\texttt{QueryAvgSalary} fails returning \texttt{null} and (2)~method
\texttt{UpdateSalary} fails returning \texttt{error}. \asd can intervene to match
their return values in successful executions, e.g., \texttt{50} and \texttt{OK},
respectively. The intervention on the first predicate does not modify any
program state and, as the successful execution shows, the return value
\texttt{50} can be safely used by the application. However, altering the return
value of \texttt{UpdateSalary}, but not updating the salary, may not be
sufficient intervention: other parts of the application that rely on the
updated salary may fail. Inferring such side-effects is hard, if not impossible.
\end{sloppypar}

\asd is restricted to safe interventions. It relies on developers to indicate
which methods do not change (internal or external) application states and
limits return-value interventions to only those methods (e.g., to
\texttt{QueryAvgSalary}, but not to \texttt{UpdateSalary}). The same holds for
exception-handling interventions. \asd removes from predicate logs
any predicates that cannot be safely intervened without undesirable
side-effects. This ensures that the rest of the \asd pipeline can safely
intervene on any subset of predicates. Excluding some interventions
may limit \asd's precision, as it may eliminate a root-cause predicate.
In such cases, \asd may find another intervenable predicate that is causally
related to the root cause, and is still useful for debugging. In our
experiments (Section~\ref{sec:experimental-evaluation}) we did not observe this issue,
since the root-cause predicates were safe to intervene.

}

\section{Approximating Causality} \label{sec:ac}
\looseness-1
\asd relies on traditional SD to derive a set of fully-discrimi\-native
predicates.
Using the logs of
successful and failed executions, \asd extracts temporal relationships among
these predicates, and uses temporal precedence to approximate causality.
\revisesix{
It is clear that in the absence of feedback loops, a cause temporally precedes
an effect~\cite{DBLP:conf/kr/PearlV91}. 
To handle loops, \asd considers multiple executions of the same program
statement (e.g., within a loop, recursion, or multiple method calls) as
separate instances, identified by their relative order of appearances during
program execution, and maps them to separate predicates~\techRepCitation. This
ensures that temporal precedence among predicates correctly
\emph{over-approximates} causality.
}

\smallskip

\noindent 
\textbf{Approximate causal DAG.} \looseness-1
\asd represents the approximation of
causality in a DAG: each node represents a predicate, and an edge $P_1
\rightarrow P_2$ indicates that $P_1$ temporally precedes $P_2$ in \emph{all}
logs where both predicates appear. Figure~\ref{illustrativeExample}(a) shows an
example of the \acmFull (\acm). 
We use circles to explicitly depict \emph{junctions} in the \acm; junctions are
not themselves predicates, but denote splits or merges in the precedence ordering of predicates. Therefore, each predicate has in- and out-degrees of at most 1, while
junctions have in- or out-degrees greater than 1.
Note that, for clarity of visuals, in our
depictions of the \acm, we omit edges implied by transitive closure. For
example, there exists an edge $P_3 \rightarrow P_5$, implied by $P_3
\rightarrow P_4$ and $P_4 \rightarrow P_5$, but it is not depicted. \asd
enforces an assumption of counterfactual causality by excluding from the \acm
any predicates that were not observed in \emph{all} failed executions: if some
executions failed without manifesting $P$, then $P$ cannot be a cause of the
failure.

\revisesix{

\smallskip
\noindent
\textbf{Completeness of \acm.} 
\looseness-1 
The \acm is complete with respect to the available, and safely-intervenable,
predicates: it contains all fully-discriminative predicates that are safe to
intervene, and if $P_1$ causes $P_2$, it includes the edge $P_1 \rightarrow
P_2$. However, it may not be complete with respect to all possible true root
causes, as a root cause may not always be represented by the available
predicates (e.g., if the true root cause is a data race and no predicate is
used to capture it). In such cases, \asd will identify the (intervenable)
predicate that is closest to the root cause and is causally related to the
failure.

Since temporal precedence among predicates is a necessary condition for
causality, the \acm is guaranteed to contain the true causal relationships.
However, temporal precedence is not sufficient for causality, and thus some
edges in the \acm may not be truly causal.
 
}

\smallskip

\noindent
\revisesix{\textbf{Temporal precedence.}
Capturing temporal precedence is not always straightforward. For simplicity of
implementation, \asd relies on computer clocks, which works reasonably well in
practice. Relying on computer clocks is not always precise as the time gap
between two events may be too small for the granularity of the clock; moreover,
events may occur on different cores or machines whose clocks are not perfectly
synchronized. These issues can be addressed with the use of logical clocks such
as Lamport's Clock~\cite{DBLP:journals/cacm/Lamport78}.}

\revisesix{Another challenge is that some predicates are associated with time windows, rather
than time points. The correct policy to resolve temporal precedence of two
temporally overlapping predicates often depends on their semantics. However,
the predicate types give important clues regarding the correct policy. In \asd,
predicate design involves specifying a set of rules that dictates the temporal
precedence of two predicates. In constructing the \acm, \asd uses those rules.

For example, consider a scenario where \texttt{foo()} calls \texttt{bar()} and
waits for \texttt{bar()} to end---so, \texttt{foo()} starts \emph{before} but
ends \emph{after} \texttt{bar()}.
\begin{itemize}
	 \item (Case 1): Consider two predicates $P_1$: ``\texttt{foo()} is running
	 slow'' and $P_2$: ``\texttt{bar()} is running slow''. Here, $P_2$ can cause
	 $P_1$ but not the other way around. In this case, \asd uses the policy that
	 \emph{end-time implies temporal precedence}.

	 \item (Case 2): Now consider $P_1$: ``\texttt{foo()} starts later than
	 expected'' and $P_2$ : ``\texttt{bar()} starts later than expected''. Here,
	 $P_1$ can cause $P_2$ but not the other way around. Therefore, in this case,
	 \emph{start-time implies temporal precedence}.
 \end{itemize} 
 
\smallskip \noindent \asd works with any policy of deciding precedence, as
long as it does not create cycles in the \acm. Since temporal precedence is a
necessary condition for causality, any conservative heuristic for deriving
temporal precedence would work. A conservative heuristic may introduce more
false positives (edges that are not truly causal), but those will be pruned by
interventions (Section~\ref{sec:refining-causality}). 

}

\begin{figure}[t]
	\centering
	\renewcommand{\arraystretch}{.95}
	{
		\setlength\tabcolsep{3pt}
		\begin{tabular}{cl}
			\toprule
			\textbf{Notation} & \multicolumn{1}{c}{\textbf{Description}}\\
			\midrule
			$\DAG$ & \acmFullFirstCap (\acm)\\
			$\causalPath$ & Causal path\\
			$F$ & Failure indicating predicate\\
			$P$ & A predicate\\
			$\Preds$ & Set of predicates\\
			$P(r)$ & Predicate $P$ is observed in execution $r$\\
			$\neg P(r)$ & Predicate $P$ is not observed in execution $r$\\
			$P_1 \leadsto P_2$ & There is a path from $P_1$ to $P_2$ in $\DAG$\\
			\bottomrule
		\end{tabular}
	}
	\vspace{0mm}
	\caption{Summary of notations used in Section~\ref{sec:refining-causality}.}
	\vspace{-0mm}
	\label{tab:notations1}
\end{figure}

\section{Causal Intervention}\label{sec:refining-causality}
\looseness-1 In this section, we describe \asd's core component, which refines
the \acm through a series of \emph{causal interventions}. An intervention on a
predicate forces the predicate to a particular state; the execution of the
application under the intervention asserts or contradicts the causal connection
of the predicate with the failure, and \asd prunes the \acm accordingly.
Interventions can be costly, as they require the application to be re-executed.
\asd minimizes this cost by (1)~smartly selecting the proper predicates to
intervene, (2)~grouping interventions that can be applied in a single
application execution, and (3)~aggressively pruning predicates even without
direct intervention, but based on outcomes of other interventions.
Figure~\ref{tab:notations1} summarizes the notations used in this section.

\looseness-1 
We start by formalizing the problem of \emph{causal path discovery} and state
our assumptions (Section~\ref{sec:prob-def-assumptions}). Then we provide an
illustrative example to show how \asd works (Section~\ref{sec:example-asd}). We
proceed to describe \emph{interventional pruning} that \asd applies to
aggressively prune predicates during group intervention rounds
(Section~\ref{predicatePruning}). Then we present \asd's causality-guided group
intervention algorithm (Section~\ref{sec:main-algo}) which administers group
interventions to derive the causal path.

\subsection{Problem Definition and Assumptions}\label{sec:prob-def-assumptions}
Given an application that intermittently fails, our goal is to provide an
informative explanation for the failure. To that end, given a set of
fully-discriminative predicates $\Preds$, we want to find an ordered subset of
$\Preds$ that defines the causal path from the root-cause predicate to the
predicate indicating the failure. Informally, \asd finds a chain of predicates
that starts from the root-cause predicate, ends at the failure predicate, and
contains the maximal number of explanation predicates such that each is caused
by the previous one in the chain. We address the problem in a similar setting
as SD, and make the following two assumptions:

\smallskip 

\noindent \emph{\textbf{Assumption 1 (Single Root-cause Predicate).}}
The root cause of a failure is the predicate whose absence (i.e., a value of
\texttt{false}) certainly avoids the failure, and there is no other predicate
that causes the root cause. We assume that in all the failed executions, there
is exactly one root-cause predicate.

\revisesix{ This assumption is prevalent in the SD
literature~\cite{statisticalDebuggingLiblit, Sober, crugLiblit}, and is
supported by several studies on real-world concurrency bug
characteristics~\cite{DBLP:conf/asplos/LuPSZ08, wong2009survey,
DBLP:conf/icsm/VahabzadehF015}, which show that a vast majority of root causes
can be captured with reasonably simple single predicates (see
Appendix). In practice, even with specific inputs, a program may fail in
multiple ways. However, failures by the same root cause generate a unique
\emph{failure signature} and hence can be grouped together using metadata
(e.g., stack trace of the failure, location of the failure in the program
binary, etc.) collected by failure
trackers~\cite{DBLP:conf/sosp/GlerumKGAONGLH09}. \asd can then treat each group
separately, targeting a single root cause for a specific failure. Moreover, the
single-root-cause assumption is reasonable in many simpler settings such as
unit tests that exercise small parts of an application.

\looseness-1 Note that this assumption does not imply that the root cause consists of a
single event; a predicate can be arbitrarily complex to capture multiple
events. For example, the predicate ``there is a data race on X'' is
\texttt{true} when two threads access the same shared memory X at the same
time, the accesses are not lock-protected, and one of the accesses is a write
operation.
Whether a single predicate is sufficient to capture the root cause
depends on predicate design, which is orthogonal to \asd.  \asd adapts
the state-of-the art predicate design, tailored to capture root causes of
concurrency bugs~\cite{crugLiblit}, which is sophisticated enough to capture
all common root causes using single predicates. If no
single predicate captures the true root cause, \asd still
finds the predicate closest to the true root cause in the true causal path.

}

\smallskip

\noindent \emph{\textbf{Assumption 2 (Deterministic Effect).}} A root-cause
predicate, if triggered, causes a fixed sequence of intermediate predicates 
(i.e., effects) before eventually causing the failure. We call this sequence
\emph{causal path}, and we assume that there is a unique one for each
root-cause-failure pair.

\revisesix{\looseness-1 Prior work has considered, 
and shown evidence of, a unique causal path between a root cause and the 
failure in sequential applications~\cite{sumner2009algorithms, 
DBLP:conf/sp/JohnsonCCMPRS11}.
The unique causal path assumption is likely to hold in concurrent applications
as well for two key reasons. First, the predicates in \asd's causal path may
remain unchanged, despite nondeterminism in the underlying instruction sequence.
For example, the predicate ``there is a data race between methods X and Y'' is
not affected by which method starts first, as long as they temporally overlap.
Second, \asd only considers fully-discriminative predicates. If such predicates
exist to capture the root cause and its effects, by the definition of being
fully discriminative, there will be a unique causal path (of predicates) from
the root cause to the failure. In all six of our real-world case studies
(Section~\ref{sec:experimental-evaluation}), such predicates existed and there
were unique causal paths from the root causes to the failures.

Note that it is possible to observe some degree of disjointness within the true
causal paths. For example, consider a case where the root cause $C$ triggers
the failure $F$ in two ways: in some failed executions, the causal path is $C
\rightarrow A_1 \rightarrow B \rightarrow F$ and, for others, $C \rightarrow
A_2 \rightarrow B \rightarrow F$. This implies that neither $A_1$ nor $A_2$ is
fully discriminative. Since \asd only considers fully-discriminative
predicates, both of them are excluded from the \acm. In this case, \asd reports
$C \rightarrow B \rightarrow F$ as the causal path; this is the shared part of
the two causal paths, which includes all counterfactual predicates and omits
any disjunctive predicates. One could potentially relax this assumption by
encoding the interaction of such predicates through a fully-discriminative
predicate (e.g., $A = A_1 \vee A_2$ encodes disjunction and is fully
discriminative).

}

\smallskip

Based on these assumptions, we define the causal path discovery problem
formally as follows.

\begin{definition}[Causal Path Discovery]\label{CPD}
Given an \acmFull $\DAG = (\Vt, \Et)$ and a predicate $F \in \Vt$
indicating a specific failure, the \textbf{causal path discovery} problem seeks a path
$\causalPath = \langle C_0, C_1, \dots, C_n \rangle$ such that the following
conditions hold:

\begin{itemize}
	\item $C_0$ is the root cause of the failure and $C_n = F$.
	\item $\forall\; 0 \le i \le n$, $C_i \in \Vt$ and
	$\forall\; 0 \le i < n$, $(C_i$, $C_{i+1}) \in \Et$.
	\item $\forall \; 0 \le i < j \le n$, $C_i$ is a counterfactual
	cause of $C_j$.
	\item $|\causalPath|$ is maximized.
\end{itemize}
\end{definition}

\begin{figure*}[t]
	\begin{center}
		\vspace{-0mm}
		\includegraphics[width=.99\textwidth]{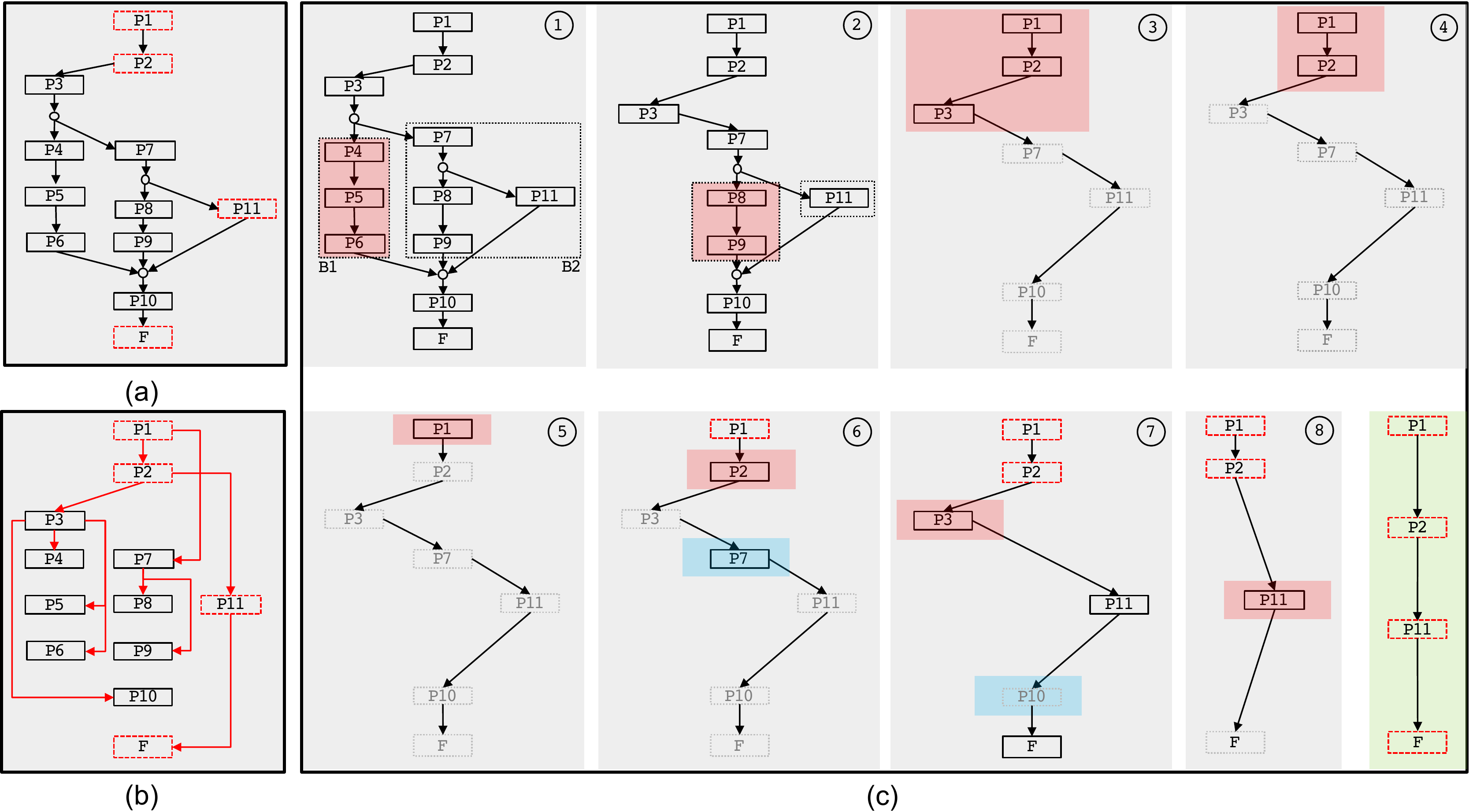}
		\vspace{-1mm}
		\caption[]{(a)~\acm as constructed by \asd. The DAG includes all
		edges implied by transitive closure, but we omit them for clarity of the visuals.
		We indicate the predicates in the causal path with the dashed red outline.
		(b)~The actual causal DAG is a subgraph of the \acm. 
		(c)~Step by step
			illustration to discover the causal path (shown at bottom right). Steps 
			\stepCounter{1} and \stepCounter{2}
			perform branch pruning, steps \stepCounter{3}--\stepCounter{8} perform 
			group intervention with pruning
			on the predicate chain, steps \stepCounter{6} and \stepCounter{7} 
			apply interventional pruning.}
		\label{illustrativeExample}
		\vspace{-0mm}
	\end{center}
\end{figure*}

\subsection{Illustrative Example}\label{sec:example-asd} 
\asd performs causal path discovery through an intervention algorithm
(Section~\ref{sec:main-algo}). Here, we illustrate the main steps and
intuitions through an example.

Figure~\ref{illustrativeExample}(a) shows an \acm derived by \asd
(Section~\ref{sec:ac}). The \acm contains all edges implied by transitive
closure, but we do not depict them to have clearer visuals. The true causal
path for the failure $F$ is $P1 \rightarrow P2 \rightarrow P11 \rightarrow F$,
depicted with dashed red outline. The \acm is a superset of the actual causal
graph, which is shown in Figure~\ref{illustrativeExample}(b).

\asd follows an intervention-centric approach for discovering the causal path.
Intervening on a predicate \emph{forces} it to behave the way it does in the
successful executions, which is by definition, the opposite of the failed
executions. (Recall that, without loss of generality, we assume that
all predicates are boolean.) Following the adaptive group testing paradigm, \asd
performs group intervention, which is simultaneous intervention on a set of
predicates, to reduce the total number of interventions.
Figure~\ref{illustrativeExample}(c) shows the steps of the intervention
algorithm, numbered \stepCounter{1}--\stepCounter{8}.

\asd first aims to reduce the \acm by pruning entire chains that are not
associated with the failure, through a process called \emph{branch pruning}
(Section~\ref{sec:main-algo}). Starting from the root of the \acm, \asd
discovers the first junction, after predicate $P3$.
For each child of
a junction, \asd creates a compound predicate, called an \emph{independent
branch}, or simply \emph{branch}, that is a disjunction over the child and all
its descendants that are not descendants of the other children. So, for the
junction after $P3$, we get branches $B1 = P4 \vee P5 \vee P6$ and $B2 = P7
\vee P8 \vee P9 \vee P11$. \asd intervenes on one of the branches chosen at
random---in this case $B1$---at step~\stepCounter{1}; this requires an
intervention on all of its disjunctive predicates ($P4$, $P5$, and $P6$) in
order to make the branch predicate \texttt{False}. Despite the intervention,
the program continues to fail, and \asd prunes the entire branch of $B1$,
resolving the junction after $P3$. For a junction of $B$ branches, \asd would
need $\log B$ interventions to resolve it using a divide-and-conquer approach.
At step \stepCounter{2}, \asd similarly prunes a branch at the junction after $P7$.
At this point, \asd is done with branch pruning since it is left with just a
chain of predicates (step~\stepCounter{3}).

What is left for \asd is to prune any non-causal predicate from the remaining
chain. \asd achieves that through a divide-and-conquer strategy that intervenes
on groups of predicates at a time (Algorithm~\ref{algo:group testing}). It intervenes on
the top half of the chain---$\{P1, P2, P3\}$---which stops the failure and
confirms that the root cause is in this group (step~\stepCounter{3}). With two
more steps that narrow down the interventions (steps~\stepCounter{4}
and~\stepCounter{5}), \asd discovers that $P1$ is the root cause. Note that we
cannot simply assume that the root of the \acm is a cause, because the edges
are not all necessarily causal.

After the discovery of the root cause, \asd needs to derive the causal path.
Continuing the divide-and-conquer steps, it intervenes on $P2$
(step~\stepCounter{6}). This stops the failure, confirming that $P2$ is in the
causal path. In addition, since $P7$ is not causally dependent on $P2$, the
intervention on $P2$ does not stop $P7$ from occurring. This observation allows
\asd to prune $P7$ without intervening on it directly. At step~\stepCounter{7},
\asd intervenes on $P3$. The effect of this intervention is that the failure is
still observed, but $P10$ no longer occurs, indicating that $P10$ is causally
connected to $P3$, but not to the failure; this allows \asd to prune both $P3$
and $P10$. Finally, at step~\stepCounter{8}, \asd intervenes on $P11$ and
confirms that it is causal, completing the causal path derivation. \asd
discovered the causal path in 8 interventions, while na\"ively we would have
needed 11---one for each predicate.

\subsection{Predicate Pruning}\label{predicatePruning} 

In the initial construction of the \acm, \asd excludes predicates based on a
simple rule: a predicate $P$ is excluded if there exists a
program execution $r$, where $P$ occurs and the failure does not
($P(r)\wedge\neg F(r)$), or $P$ does not occur and the failure does ($\neg
P(r)\wedge F(r)$). Intervening executions follow the same basic intuition for
pruning the intervened predicate $C$: By definition $C$ does not occur in an
execution $r_C$ that intervenes on predicate $C$ ($\neg C(r_C)$); thus, if the
failure still occurs on $r_C$ ($F(r_C)$), then $C$ is pruned from the \acm.

As we saw in the illustrative example, intervention on a predicate $C$ may also
lead to the pruning of additional predicates. However, the same basic pruning
logic needs to be applied more carefully in this case. In particular, we can
never prune predicates that precede $C$ in the \acm, as their potential causal
effect on the failure may be muted by the intervention on $C$. Thus, we can
only apply the pruning rule to any predicate $X$ that is not an ancestor of $C$
in the \acm ($X\not\leadsto C$). We formalize the predicate pruning strategy
over $\DAG(\Vt,\Et)$ in the following definition.

\begin{definition}[Interventional Pruning]\label{def:pruning}
	Let $R_\cSet$ be a set of program executions\footnote{\scriptsize{Because of nondeterminism issues in concurrent applications, we execute a
program multiple times with the same intervention. However, it is sufficient to
identify a single counter-example execution to invoke the pruning rule.}} intervening on a group of
predicates $\cSet\subseteq\Vt$. Every
$C\in\cSet$ is pruned from $\DAG$ iff $\exists r\in R_\cSet$
such that $F(r)$. Any other predicate $P\not\in \cSet$ is pruned from
$\DAG$ iff $\not\exists C\in\cSet$ such that $P\leadsto C$ and
$\exists r\in R_\cSet$ such that $(P(r) \wedge \neg F(r)) \vee (\neg P(r)
\wedge F(r))$.
\end{definition}

\setlength{\textfloatsep}{2pt}
\begin{algorithm}[t]
	\LinesNumbered
	\small{
		\Input{ A set of candidate predicates, $\Preds$, \\
			\acm, $\DAG$\\
			Failure indicating predicate, $F$\\
		}
		\Output{The set of counterfactual causes of $F$, $\cSet$\\
			The set of spurious predicates, $\xSet$}
		$\cSet = \emptyset$\tcc*[f]{causal predicate set}\\
		$\xSet = \emptyset$\tcc*[f]{spurious predicate set}\\
		\While{$\Preds \neq \emptyset$}
		{
			$\Preds_{1}$ = first half of $\Preds$ in topological order \label{topOrderG}\\
			$R_{\Preds_1} = $ Intervene ($\Preds_{1}$) \label{interveneG}\\
			\If(\tcc*[f]{failure stopped}){$\not\exists r \in R_{\Preds_1}$ s.t. $F(r)$\label{failStop}}
			{
				\If{$\Preds_{1}$ contains a single predicate\label{singlePredicate}}
				{
					$\cSet = \cSet \cup \Preds_{1}$ \tcc*[f]{a cause is confirmed}\label{line7}
				}
				\Else(\tcc*[f]{need to confirm causes})
				{
					$\cSet^\prime, \xSet^\prime = $ 
					\textbf{GIWP}($\Preds_{1}, \DAG, F$) \label{repeatedGIWP}\\
					$\cSet = \cSet \cup \cSet^\prime$ \tcc*[f]{confirmed causes}\\
					$\xSet = \xSet \cup \xSet^\prime$ \tcc*[f]{spurious predicates}
				}
			}
			\tcc{interventional pruning}
			\If(\tcc*[f]{failure didn't stop}){$\exists r \in R_{\Preds_1}$ s.t. $F(r)$\label{def-prune}}
			{
				$\xSet = \xSet \cup \Preds_1$ \tcc*[f]{pruning}\label{prune1}
			}

	        \ForEach{$P \in \Preds - \Preds_1$ s.t. $\forall P' \in  \Preds_1 \;\; P \not\leadsto P'$}
			{
				
				\If{$\exists r \in R_{\Preds_1}$ s.t. $(P(r) \wedge \neg F(r)) \vee (\neg P(r) \wedge F(r))$}
				{
					$\xSet = \xSet \cup \{P\}$ \tcc*[f]{pruning}\label{prune2}
				}
				
			}				
			$\Preds = \Preds - (\cSet \cup \xSet)$ \label{refineG}\tcc*[f]{remove confirmed and spurious predicates from candidate predicate pool}
		}
		\Return $\cSet$, $\xSet$\label{returnG}
	}
	\caption{GIWP ($\Preds, \DAG, F$)}
	\label{algo:group testing}
\end{algorithm}

\subsection{Causality-guided Intervention}\label{sec:main-algo}

\asd's core intervention method is described in Algorithm~\ref{algo:group testing}:
Group Intervention With Pruning (GIWP). GIWP applies adaptive group testing to
derive causal and spurious (non-causal) nodes in the \acm. The algorithm
applies a divide-and-conquer approach that groups predicates based on their
topological order (a linear ordering of its nodes such that for every directed
edge $P_1 \rightarrow P2$, $P_1$ comes before $P_2$ in the ordering).
In every iteration, GIWP selects the
predicates in the lowest half of the topological order, \revisefive{resolving ties randomly}, and intervenes by
setting all of them to \texttt{False}
(lines~\ref{topOrderG}--\ref{interveneG}).
The intervention returns a set of predicate logs.

\looseness-1
If the failure is not observed in any of the intervening executions
(line~\ref{failStop}), based on counterfactual causality, GIWP concludes that
the intervened group contains at least one predicate that causes
the failure. If the group contains a single predicate, it is marked as causal
(line~\ref{line7}). Otherwise, GIWP recurses to trace the causal predicates
within the group (line~\ref{repeatedGIWP}).

During each intervention round, GIWP applies Definition~\ref{def:pruning} to prune
predicates that are determined to be non-causal
(lines~\ref{def-prune}--\ref{prune2}). First, if the algorithm discovers an
intervening execution that still exhibits the failure, then it labels all
intervened predicates as spurious and marks them for removal (line~\ref{prune1}).
Second, GIWP examines each other predicate that does not precede any
intervened predicate and observes if any of the intervened executions
demonstrate a counterfactual violation between the predicate and the failure.
If a violation is found, that predicate is pruned (line~\ref{prune2}).

At completion of each intervention round, GIWP refines the predicate pool by
eliminating all confirmed causes and spurious predicates (line~\ref{refineG})
and enters the next intervention round . It continues the interventions until
all predicates are either marked as causal or spurious and the remaining
predicate pool is empty. Finally, GIWP returns two disjoint predicate
sets---the causal predicates and the spurious predicates (line~\ref{returnG}).

\setlength{\textfloatsep}{4pt}
\begin{algorithm}[t]
	\LinesNumbered
	\small{
		\Input{\acm, $\DAG = (\Vt, \Et)$\\
			Failure indicating predicate, $F$}
		\Output{Reduces $\DAG$ to an \acmChain}
		$\cSet = \emptyset$\tcc*[f]{potential causal predicate set}\\
		$\xSet = \emptyset$\tcc*[f]{spurious predicate set}\\
		\While{$\Vt - \cSet \neq \emptyset$}{
			$\Preds$ = predicates at the lowest topological level in $\Vt - \cSet$\label{pickTopBP}\\
			\If{$\Preds$ contains a single predicate\label{singlePath}}
			{
				$\cSet = \cSet \cup \Preds$ \tcc*[f]{add to potential causal set}
			}
			\Else(\tcc*[f]{this is a junction}\label{junction})
			{
				$\branch = \emptyset$\\
				\ForEach{$P \in \Preds$}
				{
					$\branch_{P} =  \bigvee \{Q: P \leadsto Q \wedge \forall P' \in \mathcal{P} {-} \{P\}$ $P' \not\leadsto Q\}$\label{unqiueDesc}\\
					$\branch_{P} = P \vee \branch_{P}$\\
					$\branch = \branch \cup \{\branch_{P}\} $\tcc*[f]{set of branches} \label{megaPred}
				}
				
				$\cSet', \xSet' = $ \textbf{GIWP} ($\branch, 
				\DAG, F$) \label{giwpFromBP}\\
					$\cSet = \cSet \cup \cSet'$\tcc*[f]{add to potential causal set}\\
					$\xSet = \xSet \cup \xSet'$\tcc*[f]{$\!\!$add to spurious set}\\
			}
			\tcc{refining $\DAG$}
			$\mathcal{U} = \{U: \cSet \neq \emptyset \wedge \forall C \in \cSet \; C \not\leadsto U\}$\tcc*[f]{$\!\!$unreachable$\!\!$}\\
			$\Vt = \Vt - \xSet$\tcc*[f]{remove spurious predicates}\label{removeSpurious}\\			
			$\Vt = \Vt -  \mathcal{U}$\tcc*[f]{remove unreachable predicates}\label{unreachablePreds}
		}
	}
	\caption{Branch-Prune ($\DAG, F$)}	
	\label{algo:branch-pruning}
\end{algorithm}

\setlength{\floatsep}{10pt}

\setlength{\textfloatsep}{2pt}
\DontPrintSemicolon
\begin{algorithm}[t]
	\LinesNumbered
	\small{
		\Input{\acm, $\DAG = (\Vt, \Et)$\\
			Failure indicating predicate, $F$\\
			$Flag_B$, whether to apply branch pruning or not}
		\Output{A causal path}
		\If{$Flag_B$}
		{
			\textbf{Branch-Prune} ($\DAG, F$)\label{line1}\\
		}	
		$\cSet, \xSet = $ 
		\textbf{GIWP} ($\Vt{-}\{ F\}, \DAG, F$)\label{line2}\\ 
		\Return $\cSet$\label{line3}
	}		
	\caption{Causal-Path-Discovery ($\DAG, F, Flag_B$)}	
	\label{algo:causal-path-discovery}
\end{algorithm}

\subsubsection*{Branch Pruning}

GIWP is sufficient for most practical applications and can work directly on the
\acm. However, when the \acm satisfies certain conditions (analyzed in
Section~\ref{BPTheory}), we can reduce the number of required interventions through a
process called \emph{branch pruning}. The intuition is that since there is a
single causal path that explains the failure, junctions (where multiple paths
exist) can be used to quickly identify independent branches to be pruned or
confirmed as causal as a group. The branches can be used to more effectively
identify groups for intervention, reducing the overall number of required
interventions.

\looseness-1 
Branch pruning iteratively prunes branches at junctions 
(steps~\stepCounter{1} and~\stepCounter{2} in the illustrative example) to
reduce the \acm to a chain of predicates. The process is detailed in
Algorithm~\ref{algo:branch-pruning}. The algorithm traverses the DAG based on its
topological order, and does not intervene while it encounters a single node at
a time, which means it is still in a chain (line~\ref{singlePath}). When it
encounters multiple nodes at the same topological level, it means it
encountered a junction (line~\ref{junction}). A junction means that the true causal
path can only continue in one direction, and \asd can perform group
intervention to discover it. The algorithm invokes GIWP to perform this
intervention over a set of special predicates constructed from the branches at
the encountered junction (lines~\ref{unqiueDesc}--\ref{megaPred}). A branch at
predicate $P$ is defined as a disjunctive predicate over $P$ and all
descendants of $P$ that are not descendants of any other predicate at the same
topological level as $P$. An example branch from our illustrative example is
$B1 = P4 \vee P5 \vee P6$. To intervene on a branch, one has to intervene on
all of its disjunctive predicates. The algorithm defines $\branch$ as the union
of all branches, which corresponds to a completely disconnected graph (no edges
between the nodes), thus all branch predicates are at the same topological
level. GIWP is then invoked (line~\ref{giwpFromBP}) to identify the causal branch.
The algorithm removes any predicate that is not causally connected to the
failure (line~\ref{removeSpurious}) or is no longer reachable from the correct
causal chain (line~\ref{unreachablePreds}), and updates the \acm accordingly. At
the completion of branch pruning, \asd reduces the \acm to simple chain of
predicates.

\smallskip

Finally, Algorithm~\ref{algo:causal-path-discovery} presents the overall method that
\asd uses to perform causal path discovery, which optionally invokes branch
pruning before the divide-and-conquer group intervention through GIWP.

\newcommand\Symm{Symmetric\xspace}
\newcommand\symm{symmetric\xspace}

\section{Theoretical Analysis}\label{sec:complexity-analysis}
\looseness-1
In this section, we theoretically analyze the performance of \asd in terms of
the number of interventions required to identify all {\em causal predicates},
which are the predicates causally related to the failure.\footnote{Causal
predicates correspond to \emph{faulty predicates} in group testing. This
distinction in terminology is because group testing does not meaningfully
reason about causality.} Similar to the analysis of group testing algorithms,
we study the information-theoretic lower bound, which specifies the minimum
number of bits of information that an algorithm must extract to identify all
causal predicates for any instance of a problem. We also study the lower and
the upper bounds that quantify the minimum and the maximum number of group
interventions required to identify all causal predicates, respectively, for \asd
versus a Traditional Adaptive Group Testing (TAGT) algorithm.

\looseness-1 Any group testing algorithm takes $N$ items (predicates), $D$ of
which are faulty (causal), and aims to identify all faulty items using as few
group interventions as possible. Since there are ${N \choose D}$ possible
outcomes, the information-theoretic lower bound for this problem is $\log {N
\choose D}$.
The upper bound on the number of interventions using TAGT is $\mathcal{O} (D
\log N)$, since $\log N$ group interventions are sufficient to reveal each
causal predicate. Here, we assume $D < \frac{N}{\log N}$; otherwise, a linear
approach that intervenes on one predicate at a time is preferable.

\looseness-1 We now show that the Causal Path Discovery (CPD) problem
(Definition~\ref{CPD}) can reduce the lower bound on the number of required
interventions compared to Group Testing (GT). We also show that the upper bound
on the number of interventions is lower for \asd than
TAGT, because of the two assumptions of CPD (Section~\ref{sec:prob-def-assumptions}).
In TAGT, predicates are assumed to be independent of each other, and hence,
after each intervention, decisions (about whether predicates are causal)
can be made only about the intervened predicates. In contrast, \asd uses the
precedence relationships among predicates in the \acm to (1)~aggressively
prune, by making decisions not only about the intervened predicates but also
about other predicates, and to (2)~select predicates based on the topological
order, which enables effective pruning during each intervention.

\begin{example}\label{ex:theoryExampleOne}
\looseness-1
Consider the \acm of Figure~\ref{symmetricAcm}(a), consisting of $N = 6$
predicates and the failure predicate $F$. If \asd intervenes on all
predicates in one branch (e.g., $\{A_1, B_1, C_1\}$) and finds causal
connection to the failure, it can avoid intervening on predicates in the other
branch according to the deterministic effect assumption. \asd
can also use the structure of the \acm to intervene on $A_1$ (or $A_2$)
before other predicates since the intervention can prune a large set of
predicates. Since GT algorithms do not assume relationships among predicates,
they can only intervene on predicates in random order and can make decisions
about only the intervened predicates.
\end{example}

\subsection{Search Space}

\begin{figure}[t]
	\begin{center}
		\includegraphics[width=0.45\textwidth]{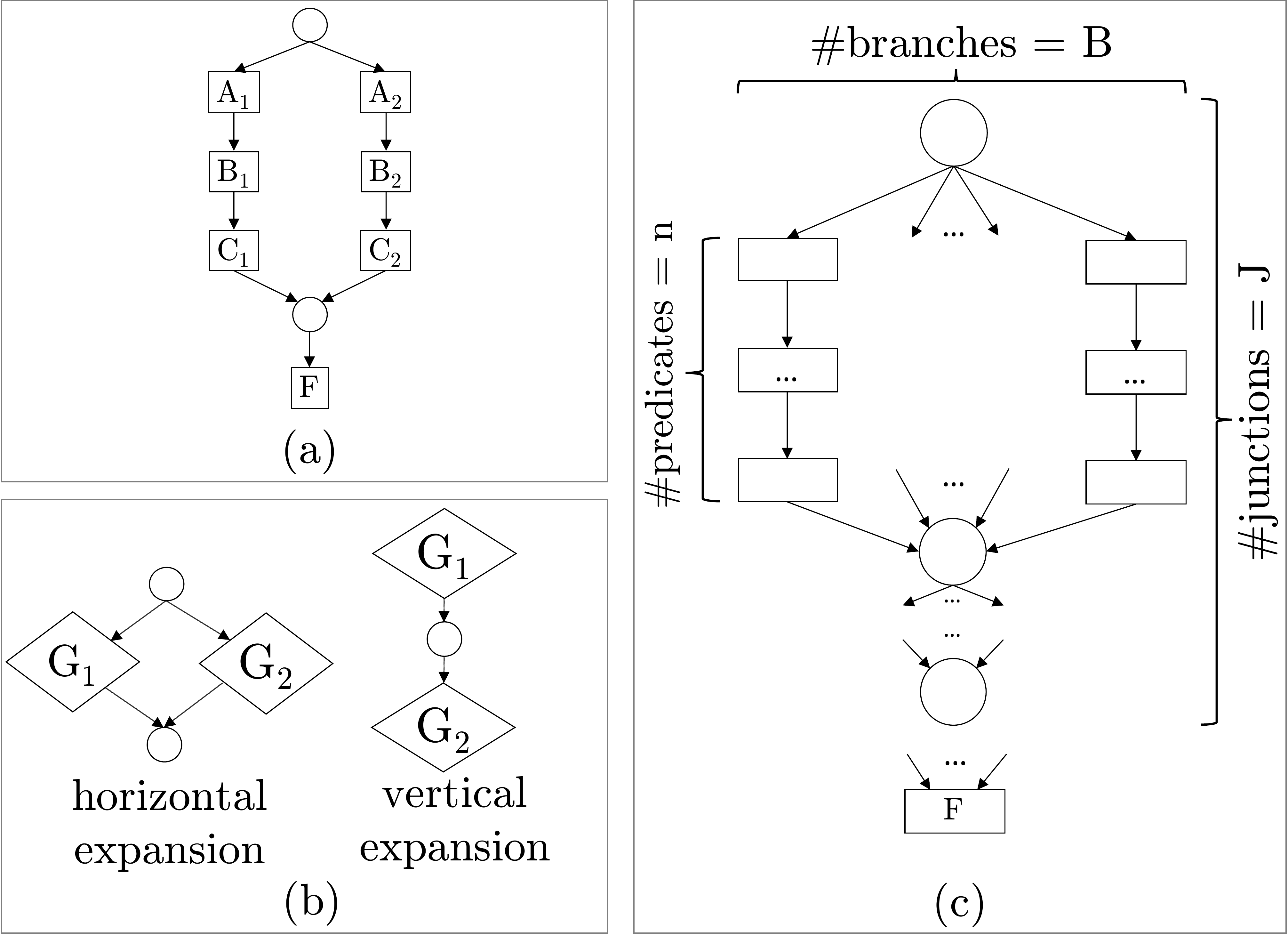}
		\vspace{-1mm}
		\caption{ 
			(a)~An \acm with failure predicate $F$.
			(b)~Horizontal and vertical expansion.
			(c)~A \symm \acm with $J$ junctions where each junction has $B$ branches and each branch has $n$ predicates.}
		\vspace{1mm}
		\label{symmetricAcm}
	\end{center}
\end{figure}

The temporal precedence and potential causality encoded in the
\acm restrict the possible causal paths and significantly reduce the search
space of CPD compared to GT. 
\begin{example}\label{ex:oneBubble}
    In the example of Figure~\ref{symmetricAcm}(a), GT considers all subsets of the 6
    predicates as possible solutions, and thus its search space includes $2^6 =
    64$ candidates. CPD leverages the \acm and the deterministic effect
    assumption (Section~\ref{sec:prob-def-assumptions}) to identify invalid candidates
    and reduce the search space considerably. For example, the candidate
    solution $\{A_1, B_2, C_1\}$ is not possible under CPD, because it involves
    predicates in separate paths on the \acm. In fact, based on the \acm, CPD
    does not need to explore any solutions with more than 3 predicates. The
    complete search space of CPD includes all subsets of predicates along each
    branch of length 3, thus a total of $2\cdot(2^3 - 1) + 1 = 15$ possible
    solutions.
\end{example}

\looseness-1
We proceed to characterize the search space of CPD compared to GT more
generally. We use $|G|$ to denote the number of predicates in an \acm
represented by $G$, and $W_G^{GT}$ and $W_G^{CPD}$ to denote the size of
the search space for GT and CPD, respectively. We start from the simplest case
of DAG, a chain, and then using the notions of horizontal and vertical
expansion, we can derive the search space for any DAG.

If $G$ is a simple chain of predicates, then GT and CPD have the same search
space: $2^{|G|}$. CPD reduces the search space drastically when junctions split
the predicates into separate branches, like in Example~\ref{ex:oneBubble}. We call
this case a \emph{horizontal expansion}: a DAG $\mathcal{G}_H$ is a horizontal
expansion of two subgraphs $G_1$ and $G_2$ if it connects them in parallel
through two junctions, at the roots (lowest topological level) and
leaves (highest topological level). In contrast, $\mathcal{G}_V$ is a
\emph{vertical expansion}, if it connects them sequentially via a junction.
Horizontal and vertical expansion are depicted in Figure~\ref{symmetricAcm}(b). In
horizontal expansion, the search space of CPD is additive over the combined
DAGs, while in vertical expansion it is multiplicative.

\begin{lemma}[DAG expansion]\label{lem:expansion}
    Let $W_{G_1}^{CPD}$ and $W_{G_2}^{CPD}$ be the numbers of valid solutions
    for CPD over DAGs $G_1$ and $G_2$, respectively. Let $\mathcal{G}_H$ and
    $\mathcal{G}_V$ represent their horizontal and vertical expansion,
    respectively. Then:
\begin{align*}
    W^{CPD}_{\mathcal{G}_H} &= 1 + (W_{G_1}^{CPD} - 1) + (W_{G_2}^{CPD} - 1)\\
    W^{CPD}_{\mathcal{G}_V} &= W_{G_1}^{CPD}W_{G_2}^{CPD}
\end{align*}  
In contrast, in both cases, the search space of GT is $2^{|G_1|+|G_2|}$.  
\end{lemma}

Intuitively, in horizontal expansion, the valid solutions for $\mathcal{G}_H$
are those of $G_1$ and those from $G_2$, but no combinations between the two
are possible. Note that both $W_{G_1}^{CPD}$ and $W_{G_2}^{CPD}$ have the empty
set as a common solution, so in the computation of $W^{CPD}_{\mathcal{G}_H}$,
one solution is subtracted from each search space ($W_{G_i}^{CPD} - 1$) and
then added to the overall result.

\smallskip
\noindent
\textbf{\Symm \acm.}
Lemma~\ref{lem:expansion} allows us to derive the size of the search space for CPD
over any \acm. To further highlight the difference between GT and CPD, we
analyze their search space over a special type of \acm, a \symm \acm, depicted
in Figure~\ref{symmetricAcm}(c). A \symm \acm has $J$ junctions, and $B$ branches at
each junction, where each branch is a simple chain of $n$ predicates.
Therefore, the total number of predicates in the \symm \acm is $N = JBn$, and
the search space of GT is $W^{GT} = 2^{JBn}$. For CPD, based on horizontal
expansion, the subgraph in-between two subsequent junctions has a total of
$1+\sum_i^B(2^n-1)=1+B(2^n-1)$ candidate solutions. Then, based on vertical
expansion, the overall search space of CPD is:
\[
W^{CPD} = (B(2^n - 1) + 1)^J
\] 

\smallskip

\subsection{Lower Bound of Number of Interventions} 
\looseness-1 We now show that, due to the predicate pruning mechanisms, and the
strategy of picking predicates according to topological order, the lower
bound\footnote{Lower bound is a theoretical bound which states that, it might
be possible to design an algorithm that can solve the problem which requires
number of steps equal to the lower bound. Note that, this does not imply that
there exists one such algorithm.} on the required number of interventions in
CPD is significantly reduced. For the sake of simplicity, we drop the deterministic effect assumption in this analysis. In GT,
after each group test, we get at least $1$ bit of information. Since after
retrieving all information, the remaining information should be $ \le 0$,
therefore, the number of required interventions in GT is bounded below by $\log
{N \choose D}$. In contrast, for CPD, we have the following theorem. (Proofs
are in \appOrTechRep.)

\begin{theorem}\label{THEM2}
	 The number of required group interventions in CPD is bounded below by
	 $\frac{N}{N + DS_1}\log{N \choose D}$, where at least $S_1$ predicates are
	 discarded (either pruned using the pruning rule or marked as causal) during
	 each group intervention.
\end{theorem}

Since $\frac{DS_1}{N} > 0$, we obtain a reduced lower bound for the number of
required interventions in CPD than GT. In general, as $S_1$ increases, the
lower bound in CPD decreases. Note that we are not claiming that \asd achieves
this lower bound for CPD; but this sets the possibility that improved
algorithms can be designed in the setting of CPD than GT.

\smallskip

\noindent\textbf{\Symm \acm.} Figure~\ref{tab:theory-summary} shows the lower
bound on the number of required interventions in CPD and GT for the \symm \acm
of Figure~\ref{symmetricAcm}(c), assuming that each intervention discards at
least $S_1$ predicates in CPD.

\subsection{Upper Bound of Number of Interventions}
We now analyze the upper bound on the number of interventions for \asd under
(1)~branch pruning, which exploits the deterministic effect assumption, and
(2)~predicate pruning.

\setlength{\tabcolsep}{3pt}
\begin{figure}[tb]
	\vspace{0mm}
        \renewcommand{\arraystretch}{1.4}
    \resizebox{\linewidth}{!}{
	{\small
	\begin{tabular}{|l|c|c|c|}
		\hline
		& Search & \multicolumn{2}{c|}{\#Interventions}\\
		\cline{3-4}
		& space & Lower bound & Upper bound (AID/TAGT)\\
		\hline
		\hline
		CPD
		&  $(B(2^n{-}1){+} 1)^J$ 
		& \hspace{0.2mm}$\frac{JBn}{JBn + DS_1} \log {JBn \choose D}$  
		& $J \log B {+} D \log {(Jn)} - \frac{D(D-1)S_2}{2Jn}$\\[1pt]
		
		\hline
		
		GT 
		& $2^{JBn}$ 
		& $\log  {JBn \choose D}$ 
		& $D \log B {+} D \log {(Jn)} - \frac{D(D-1)}{2JBn}$ \\[1pt]
		
		\hline
	\end{tabular}
	}}
	\vspace{0mm}
	\caption{Theoretical comparison between CPD and GT
	for the \symm \acm of Figure~\ref{symmetricAcm}(c).}
	\vspace{6mm}
	\label{tab:theory-summary}
\end{figure}
\setlength{\tabcolsep}{6pt}

\subsubsection{Branch Pruning}\label{BPTheory} \looseness-1 
Whenever \asd encounters a junction, it has the option to apply branch pruning.
In CPD, at most one branch can be causal at each junction; hence, we can find
the causal branch using $\log B$ interventions at each junction, where $B$ is the number of
branches at that junction. Also, $B$ is upper-bounded by the number of threads $T$
in the program. This holds since we assume that the program inputs are fixed
and there is no different conditional branching due to input variation in
different failed executions within the same thread.
If there are $J$ junctions and at most $T$ branches at each junction, the
number of interventions required to reduce the \acm to a chain is at most
$J\log T$. Now let us assume that the maximum number of predicates in any path
in the \acm is ${N_{M}}$. Therefore, the chain found after branch pruning can
contain at most ${N_{M}}$ predicates. If $D$ of them are causal predicates, we
need at most $D \log {N_{M}}$ interventions to find them. Therefore, the total
number of required interventions for \asd is $\le J \log T + D \log {N_{M}}$. In
contrast, the number of required interventions for TAGT, which does not prune
branches, is $ \le D \log (T{N_{M}}) = D \log T + D \log {N_{M}}$. Therefore,
whenever $J < D$, the upper bound on the number of interventions for \asd is
smaller than the upper bound for TAGT.

\subsubsection{Predicate Pruning} 

For an \acm with $N$ predicates, $D$ of which
are causal, we now focus on the upper bound on the number of interventions in
\asd using only predicate pruning. In the worst case, when no pruning is
possible, the number of required interventions would be the same as that of
TAGT without pruning, i.e., $\mathcal{O}(D\log N)$.

\begin{theorem} \label{THEM4}	
	 If at least $S_2$ predicates are discarded (pruned or marked as causal)
	 from the candidate predicate pool during each causal predicate discovery,
	 then the number of required interventions for \asd is $\le D \log N -
	 \frac{D(D-1)S_2}{2N}$.
\end{theorem}

Hence, the reduction depends on $S_2$. When $S_2 = 1$, we are referring to
TAGT, in absence of pruning, because once TAGT finds a causal predicate, it
removes that predicate from the candidate predicate pool.

\smallskip

\noindent\textbf{\Symm \acm.} Figure~\ref{tab:theory-summary} shows the upper
bound on the number of required interventions using \asd and TAGT for the \symm
\acm of Figure~\ref{symmetricAcm}(c), assuming that at least $S_2$ predicates
are discarded during each causal predicate discovery by \asd.

\section{Experimental Evaluation}\label{sec:experimental-evaluation}
\looseness-1
We now empirically evaluate \asd. We first use \asd on six real-world
applications to demonstrate its effectiveness in identifying root cause and
generating explanation on how the root cause causes the failure. Then we use a
synthetic benchmark to compare \asd and its variants against traditional
adaptive group testing approach to do a sensitivity analysis of \asd on various
parameters of the benchmark.

\subsection{Case Studies of Real-world Applications} \looseness-1 We now use three
real-world open-source applications and three proprietary applications to
demonstrate \asd's effectiveness in identifying root causes of transient
failures. Figure~\ref{tab:casestudies} summarizes the results and highlights
the key benefits of \asd:
\begin{itemize}
	
	 \item \asd is able to identify the true root cause and generate an
	 explanation that is consistent with the explanation provided by the
	 developers in corresponding GitHub issues.
	
	 \item \asd requires significantly fewer interventions than traditional
	 adaptive group testing (TAGT), which does not utilize causality among
	 predicates (columns 5 and 6).
	
	 \item In contrast, SD generates a large number of discriminative
	 predicates (column 3), only a small number of which is actually causally
	 related to the failures (column 4). 
 
 \end{itemize}

\subsubsection{Data race in Npgsql}
As a case study, we first consider a recently discovered concurrency bug in
Npgsql~\cite{npgsql}, an open-source ADO.NET Data Provider for PostgreSQL.
The bug (GitHub issue \#2485) causes an Npgsql-baked application to 
intermittently crash when it tries to create a new PostgreSQL
connection. We use \asd to check if it can identify the root cause and
generate an explanation of how the root cause triggers the failure.

We used one of the existing unit tests in Npgsql that causes the issue, and
generated logs from 50 successful executions and 50 failed executions of the
test. By applying SD, we found a total of 14 discriminative
predicates.
However, SD did not pinpoint the root cause or generate any explanation.
 
We then applied \asd on the discriminative predicates. In the branch
pruning step, it used 3 rounds of interventions to prune 8 of the 14
predicates. 
In the next step, it required 2 more rounds of interventions. Overall, \asd
required a total of 5 intervention rounds; in contrast, TAGT would require 11
interventions in the worst case.

\looseness-1
After all the interventions, \asd identified a data race as the root cause of
the failure and produced the following explanation: (1) two threads race on an
index variable: one increments it while the other reads it (2) The second
thread accesses an array at the incremented index location, which is outside
the array size. (3) This access throws \texttt{IndexOutOfRange} exception (4)
Application fails to handle the exception and crashes. This explanation matches
the root cause provided by the developer who reported the bug to Npgsql GitHub
repository.

\begin{figure}[t]
	\centering
	\renewcommand{\arraystretch}{1.2}
	\resizebox{1\linewidth}{!}{\small
		\setlength\tabcolsep{1.5pt}
	\begin{tabular}{|l|p{1.2cm}||p{1.4cm}|p{1.4cm}||c|c|}
		\hline
		\multicolumn{1}{|c|}{(1)} & \multicolumn{1}{c||}{(2)} & \multicolumn{1}{c|}{(3)} & \multicolumn{1}{c||}{(4)}  & \multicolumn{2}{c|}{\#Interventions} \\
		\cline{5-6} 
		\multicolumn{1}{|c|}{\shortstack{Application\\ \phantom{nothing}}} & 
		\multicolumn{1}{c||}{\shortstack{GitHub \\ Issue \#}}   &  
		\multicolumn{1}{c|}{\shortstack{\#Discrim. \\ preds (SD)}} & 
		\multicolumn{1}{c||}{\shortstack{\#Preds~in \\ causal~path}} &
		\multicolumn{1}{c|}{\shortstack{\\(5)\\\phantom{i}\asd\phantom{i}}}&
		\multicolumn{1}{c|}{\shortstack{(6)\\TAGT}}\\
		\hline
		\hline
		Npgsql~\cite{npgsql} & 2485~\cite{npgsqlBug2485}  & 
		\multicolumn{1}{c}{14} & \multicolumn{1}{|c||}{3} & \multicolumn{1}{c|}{5} & \multicolumn{1}{c|}{11}\\
		\hline
		Kafka~\cite{kafka} & 279~\cite{kafkaBug279} & 
		\multicolumn{1}{c}{72} & \multicolumn{1}{|c||}{5} & \multicolumn{1}{c|}{17} & \multicolumn{1}{c|}{33}\\
		\hline
		Azure Cosmos DB~\cite{cosmosdb} & 713~\cite{cosmosBug713} & 
		\multicolumn{1}{c}{64} & \multicolumn{1}{|c||}{7} &  \multicolumn{1}{c|}{15} & \multicolumn{1}{c|}{42}\\
		\hline
		\textcolor{\red}{{\tt Network}} & 
		\textcolor{\red}{N/A} & 
		\multicolumn{1}{c}{\textcolor{\red}{24}} & 
		\multicolumn{1}{|c||}{\textcolor{\red}{1}} &  
		\multicolumn{1}{c|}{\textcolor{\red}{2}} & 
		\multicolumn{1}{c|}{\textcolor{\red}{5}}\\ 
		\hline
		\textcolor{\red}{{\tt BuildAndTest}} & 
		\textcolor{\red}{N/A} & 
		\multicolumn{1}{c}{\textcolor{\red}{25}} & 
		\multicolumn{1}{|c||}{\textcolor{\red}{3}} &  
		\multicolumn{1}{c|}{\textcolor{\red}{10}} & 
		\multicolumn{1}{c|}{\textcolor{\red}{15}}\\ 
		\hline
		\textcolor{\red}{{\tt HealthTelemetry}} & 
		\textcolor{\red}{N/A} & 
		\multicolumn{1}{c}{\textcolor{\red}{93}} & 
		\multicolumn{1}{|c||}{\textcolor{\red}{10}} &  
		\multicolumn{1}{c|}{\textcolor{\red}{40}} & 
		\multicolumn{1}{c|}{\textcolor{\red}{70}}\\ 
		\hline
	\end{tabular}
	}
	 \caption{Results from case studies of real-world applications. SD
	 produces way too many spurious predicates beyond the correct causal
	 predicates (columns 3 \& 4). SD actually produces even more predicates, but
	 here we only report the number of fully-discriminative predicates. \asd and
	 traditional adaptive group testing (TAGT) both pin-point the correct causal predicates using
	 interventions, but \asd does so with significantly fewer interventions
	 (columns 5 \& 6).}
	 \vspace{4mm}
	\label{tab:casestudies}
\end{figure}

\subsubsection{Use-after-free in Kafka}
Next, we use \asd on an application built on Kafka~\cite{kafka}, a distributed
message queue. On Kafka's GitHub repository, a user reported an
issue~\cite{kafkaBug279} that causes a Kafka application to intermittently
crash or hang. The user also provided a sample code to reproduce the issue; we
use a similar code for this case study.

As before, we collected predicate logs from 50 successful and 50 failed
executions. Using SD, we identified 72 discriminative predicates. The \acm
identified 30 predicates with no causal path to the failure indicating
predicate, and hence were discarded. \asd then used the intervention algorithm
on the remaining 42 predicates. After a sequence of 7 interventions, \asd could
identify the root-cause predicate. It took an additional 10 rounds (total 17)
of interventions to discover a causal path of 5 predicates that connects the
root cause and the failure. The causal path gives the following explanation:
(1) The main thread that creates a Kafka \texttt{consumer} $C$ starts a child
thread (2) the child thread runs too slow before calling a method on $C$ (3)
main thread disposes $C$ (4) child thread calls a commit method on $C$ (5)
since $C$ has already been disposed by the main thread, the previous step
causes an exception, causing the failure. The explanation matches well with the
description provided in GitHub.

Overall, \asd required 17 interventions to discover the root cause and
explanation. In contrast, SD generates 72 predicates, without pinpointing the
true root cause or explanation. TAGT could identify all predicates in the
explanation, but it takes 33 interventions in the worst case.

\subsubsection{Timing bug in Azure Cosmos DB application} Next, we use \asd
on an application built on Azure Cosmos DB~\cite{cosmosdb}, Microsoft's
globally distributed database service for operational and analytics workloads.
The application has an intermittent timing bug similar to the one mentioned in
a Cosmos DB's pull request on GitHub~\cite{cosmosBug713}. In summary, the
application populates a cache with several entries that would expire after 1
second, performs a few tasks, and then accesses one of the cached entries.
During successful executions, the tasks run fast and end before the cached
entries expire. However, a transient fault triggers expensive fault handling
code that makes a task run longer than the cache expiry time. This makes
the application fail as it cannot find the entry in the cache (i.e., it has
already expired).

Using SD, we identified 64 discriminative predicates from successful and failed
executions of the application. Applying \asd on them required 15 interventions
and it generated an explanation consisting of 7 predicates that are consistent
with the aforementioned informal explanation. In contrast, SD would generate 64
predicates and TAGT would take 42 interventions in the worst case.

\setlength{\tabcolsep}{6pt} 
\begin{figure}[t]
	\begin{center}
	\includegraphics[width=.48\textwidth, trim=0 0 0 0 0]{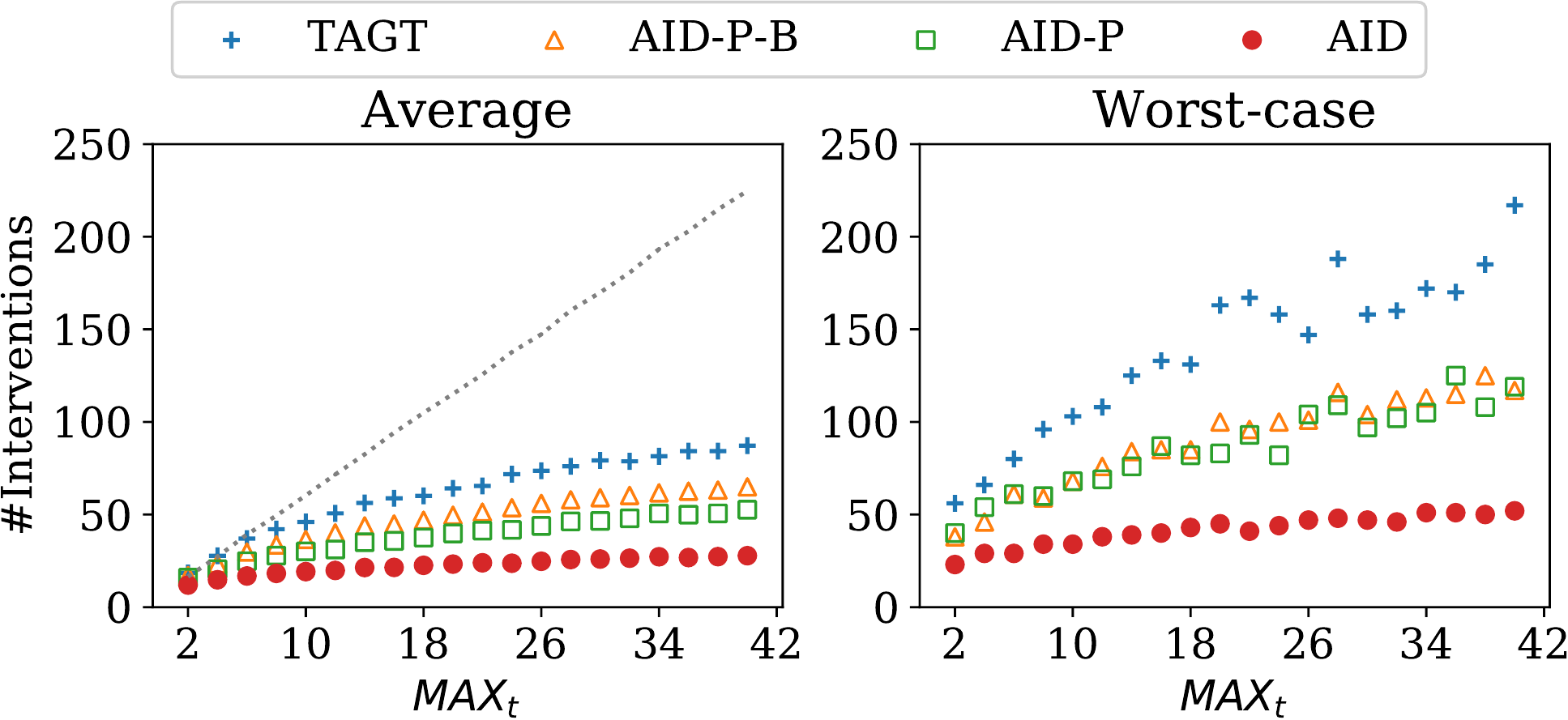}
	\vspace{-5mm}
	 \caption{Number of interventions required in the average and worst
	 case by traditional adaptive group testing (TAGT) and different variations of
	 \asd with varying $MAX_t$. For average case analysis, total number of
	 predicates is shown using a grey dotted line. Total number of predicates is
	 not shown for the worst-case analysis, because the worst cases vary across
	 approaches.}
	 \vspace{2mm}
	\label{allPlot}
	\end{center}
\end{figure}

\subsubsection{Bugs in proprietary software}\label{new-results}
\looseness-1
We applied \asd for finding root causes of intermittent failures of several 
proprietary applications inside Microsoft. We here report our experience with 
three of the applications that we name as follows 
(Figure~\ref{tab:casestudies}):
(1)~{\tt Network}: the control plane of a data center network, 
(2)~{\tt BuildAndTest}: a large-scale software build and test platform, and
(3)~{\tt HealthTelemetry}: a module used by various services to report their 
runtime health.
Parts of these applications (and associated tests) had been intermittently 
failing for several months and their developers could not identify the exact 
root causes. This highlights that the root causes of these failures were 
non-trivial. \asd identified the root causes and generated explanations for how 
the root causes lead to failures: for {\tt Network}, the root cause was a 
random number collision, for {\tt BuildAndTest}, it was an order violation of 
two events, and for {\tt HealthTelemetry}, it was a race condition. Developers 
of the applications confirmed that the root causes identified by \asd are 
indeed the correct ones and that the explanations given by \asd correctly 
showed how the root causes lead to the (intermittent) failures.

Figure~\ref{tab:casestudies} also shows the performance of \asd with these 
applications. As before, SD produces many discriminative predicates, only a 
subset of which are causally related to the failures. Moreover, for all 
applications, \asd requires significantly fewer interventions than what TAGT 
would require in the worse case.

\subsection{Sensitivity Analysis}

We further evaluate \asd on a benchmark of synthetically-generated
applications, designed to fail intermittently and with known root causes. We
generate multi-threaded applications ranging the maximum number of threads
$MAX_t$ from 2 to 40. For each parameter setting, we generate $500$
applications. In these applications, the total number of predicates $N$ ranges
from 4 to 284, and we randomly choose the number of causal predicates in the
range $[1, \frac{N}{\log N}]$.

For this experiment, we compare four approaches: TAGT, \asd, \asd without
predicate pruning (\asd-P), and \asd without predicate or branch pruning
(\asd-P-B). All four approaches derive the correct causal paths but differ in
the number of required interventions. Figure~\ref{allPlot} shows the average
(left) and the maximum (right) number of interventions required by each
approach. The grey dotted line in the average case shows the average number of
predicates over the 500 instances for that setting. This experiment provides
two key observations:

\smallskip

\noindent 
\textbf{Interventions in topological order converge faster.}
Causally-related predicates are likely to be topologically close to each other
in the \acm. \asd discards all predicates in an intervened group only when none
are causal. This is unlikely to occur when predicates are grouped randomly. For
this reason, \asd-P-B, which uses topological ordering, requires fewer
interventions than TAGT.

\smallskip

\noindent 
\textbf{Pruning reduces the required number of interventions.}
\looseness-1
We observe that both predicate and branch pruning reduce the number of
interventions. Pruning is a key differentiating factor of \asd from TAGT. In
the worst-case setting in particular, the margin between \asd and TAGT is
significant: TAGT requires up to 217 interventions in one case, while the
highest number of interventions for \asd is 52.

\section{Related Work}\label{sec:related-work}
\looseness-1
\textbf{Causal inference} has been long applied for root-cause analysis of
program failures. Attariyan et al.~\cite{DBLP:conf/osdi/AttariyanCF12,
DBLP:conf/osdi/AttariyanF10}
observe causality within application components through runtime control and
data flow; but only report a list of root causes ordered by the likelihood of
being faulty, without providing further causal connection between root causes
and performance anomalies. 
\nocite{sumner2009algorithms}
Beyond statistical association (e.g., correlation) between root cause and
failure, few techniques~\cite{Baah:2010:CIS:1831708.1831717,
DBLP:conf/sigsoft/BaahPH11, DBLP:conf/icst/ShuSPC13,
DBLP:journals/corr/abs-1712-03361} apply statistical causal inference on
observational data towards software fault localization.
\revisesix{However, observational data collected from program execution logs is
often limited in capturing certain scenarios, and hence, observational study is
ill-equipped to identify the intermediate explanation predicates. This is
because observational data is not generated by randomized controlled
experiments, and therefore,
may not satisfy
\emph{conditional exchangeability} (data can be treated as if they came from a
randomized experiment~\cite{DBLP:conf/uai/JensenBR19})
and \emph{positivity} (all possible combinations of values for the variables
are observed in the data)---two key requirements for applying causal
inference on observational data~\cite{DBLP:conf/icst/ShuSPC13}. While
observational studies are extremely useful in many settings, \asd's problem
setting permits interventional studies, which offer increased reliability and
accuracy.}

\revisesix{ \textbf{Explanation}-centric approaches are relevant to \asd as
they also aim at generating informative, yet minimal, explanations of certain
incidents, such as data errors~\cite{DBLP:conf/sigmod/WangDM15} and binary
outcomes~\cite{DBLP:journals/pvldb/GebalyAGKS14}, however these do not focus on
interventions. Viska~\cite{DBLP:conf/sigmod/GudmundsdottirS17} allows the users
to perform intervention on system parameters to understand the underlying
causes for performance differences across different systems. None of these
systems are applicable for finding causally connected paths that explain
intermittent failures due to concurrency bugs. }

\textbf{Statistical debugging} approaches~\cite{DBLP:conf/icse/ChilimbiLMNV09,
crugLiblit, statisticalDebuggingLiblit, Sober, DBLP:conf/issta/ThakurSLL09,
DBLP:conf/icml/ZhengJLNA06, DBLP:conf/sosp/KasikciCGN17,
DBLP:conf/sosp/KasikciCGN17} employ statistical diagnosis to rank 
program predicates based on their likelihood of being the root causes of 
program failures. 
However, all statistical debugging approaches suffer from the issue of not
separating correlated predicates from the causal ones, and fail to provide
contextual information regarding \emph{how} the root causes lead to program
failures. 

\reviseseven{\textbf{Predicates} in \asd are extracted from execution 
traces of the application. Ball et al.~\cite{DBLP:journals/toplas/BallL94} 
provide algorithms for  efficiently tracing execution with minimal 
instrumentation. While the authors  had a different goal (i.e., path profiling) 
than ours, the traces can be used to extract \asd predicates.}

\looseness-1 \textbf{Fault injection}
techniques~\cite{DBLP:conf/sigmod/AlvaroRH15,
han1995doctor,DBLP:journals/tc/KanawatiKA95, marinescu2009lfi} intervene
application runtime behavior with the goal to test if an application can handle
the injected faults. In fault injection techniques, faults to be injected are
chosen based on whether they can occur in practice. In contrast, \asd
intervenes with the goal of verifying (presence or absence of) causal
relationship among runtime predicates, and faults are chosen based on if they
can alter selected predicates.

\textbf{Group testing}~\cite{DBLP:conf/isit/AgarwalJM18,
DBLP:journals/sigpro/BaiWLLLZ19, hwang-group-testing,
DBLP:conf/isit/BaldassiniJA13, DBLP:conf/isit/LiCHKJ14,
dingzhu1993combinatorial, DBLP:conf/itw/KarbasiZ12} has been applied for fault
diagnosis in prior literature~\cite{DBLP:conf/uai/ZhengRB05}. Specifically,
adaptive group testing is related to \asd's intervention algorithm. However,
none of the existing works considers the scenario where a group test might
reveal additional information and thus offers an inefficient solution for
causal path discovery.

\textbf{Control flow graph}-based techniques~\cite{ChengLZWY09,
contextAwareControlFlowPath} aim at identifying bug signature for sequential
programs using discriminative subgraphs within the program's control flow
graph; or generating faulty control flow paths that link many bug predictors.
But these approaches do not consider causal connection among these bug
predictors and program failure. 

\bigskip

\looseness-1\textbf{Differential slicing}~\cite{DBLP:conf/sp/JohnsonCCMPRS11}
aims towards discovering causal path of execution differences but requires
complete program execution trace generated by execution
indexing~\cite{DBLP:conf/pldi/XinSZ08}. Dual
slicing~\cite{DBLP:conf/issta/WeeratungeZSJ10} is another program slicing-based
technique to discover statement level causal paths for concurrent program
failures. However, this approach does not consider compound predicates that
capture certain runtime conditions observed in concurrent programs. Moreover,
program slicing-based approaches cannot deal with a set of executions, instead
they only consider two executions---one successful and one failed.

\section{Conclusions}\label{sec:conclusions}
In this work, we defined the problem of causal path discovery for explaining
failure of concurrent programs. Our key contribution is the novel Adaptive
Interventional Debugging (\asd) framework, which combines existing statistical
debugging, causal analysis, fault injection, and group testing techniques in a
novel way to discover root cause of program failure and generate the causal
path that explains how the root cause triggers the failure. Such explanation
provides better interpretability for understanding and analyzing the root
causes of program failures. We showed both theoretically and empirically that
\asd is both efficient and effective to solve the causal path discovery
problem. As a future direction, we plan to incorporate additional information
regarding the program behavior to better approximate the causal relationship
among predicates, and address the cases of multiple root causes and multiple
causal paths. Furthermore, we plan to address the challenge of explaining
multiple types of failures as well.

\balance
\bibliographystyle{abbrv}
{\small
\bibliography{aid}
}

\appendix
\section{Program Instrumentation}\label{ap:instrum}
\asd separates program instrumentation and predicate extraction unlike prior SD
techniques~\cite{statisticalDebuggingLiblit, crugLiblit, Sober}. One advantage
of our separation of instrumentation and predicate extraction is that it
enables us to design predicates after collection of the application's execution
traces. In contrast, prior works in SD instrument applications to directly
extract the predicates. For example, to assess if two methods return the same
value, prior work would instrument the program using a hard coded conditional
statement ``\texttt{pred = (foo() == bar())}''. In contrast, our
instrumentation simply collects the return values of the two methods and stores
them in the execution trace. \asd later evaluates the predicates based on the
execution traces. This gives us the flexibility to design predicates
post-execution, often based on knowledge of some domain-expert. For example, in
this case, we can design multiple predicates such as whether two values are
equal, unequal, or satisfy any custom relation.

\smallskip

\noindent\textbf{Instrumentation granularity.} Instrumentation granularity is
orthogonal to \asd. Like prior SD work, we could have instrumented at a finer
granularity such as at each conditional branch; but instrumenting method calls
were sufficient for our purpose. Since our instrumentation is of much sparser
granularity than existing SD work~\cite{statisticalDebuggingLiblit, crugLiblit,
Sober} that employ sampling based finer granularity instrumentation, we do not
use any sampling.

\section{Predicate Extraction and Fault Injection}\label{predExtFIDetails}
\begin{figure}[t]
	\begin{center}
		\includegraphics[width=.46\textwidth, trim=30 0 0 0]{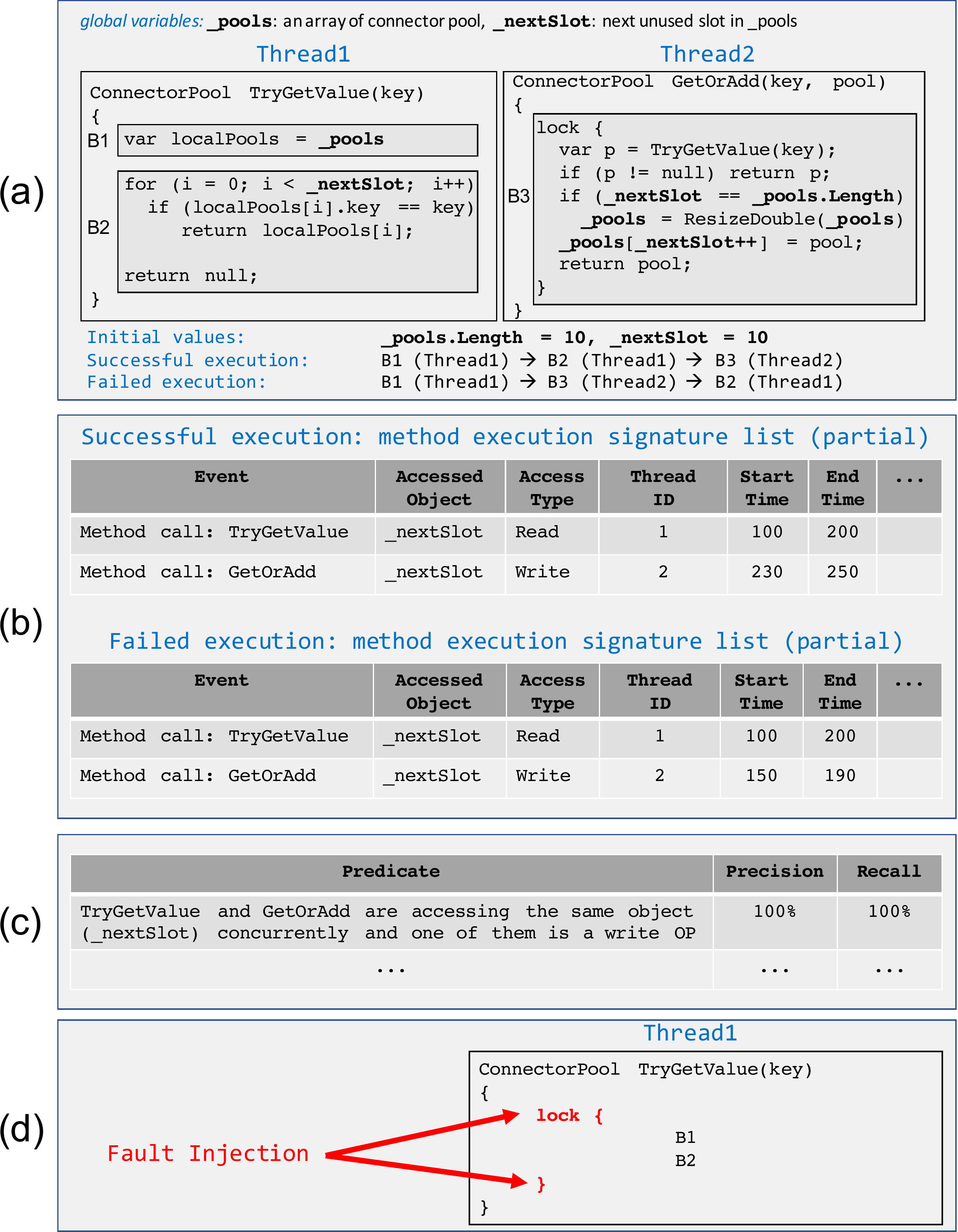}
		\vspace{5mm}
		\caption[]{ (a)~Simplified code for the Npgsql bug of Example~\ref{ex:npgsql}.
		(b)~Partial execution traces of one successful and one failed execution. 
			The start-time and end-time of the events reflect concurrent read/write access to the shared variable \texttt{\_nextSlot}.
		(c)~The race predicate is one of the discriminative predicates. 
		(d)~Fault injection to intervene (disable) the race predicate: putting a lock around the instructions within \texttt{TryGetValue()}.}
		\label{fig:runningExample}
		\vspace{4mm}
	\end{center}
\end{figure}

Figure~\ref{fig:runningExample} shows the complete pipeline of predicate
extraction and fault-injection for the Npgsql bug of Example~\ref{ex:npgsql},
whose simplified source code is shown in Figure~\ref{fig:runningExample}(a).
Executions of the instrumented application generate a list of runtime method
signatures per execution, called \emph{execution traces}. Two partial execution
traces---one for a successful and the other for a failed execution---are shown
in Figure~\ref{fig:runningExample}(b). Then we extract predicates and compute
their precision and recall as shown in Figure~\ref{fig:runningExample}(c).

In \asd, we use existing fault injection techniques---which is able to change a
method's input and return value, can cause a method to throw exception, can
cause a method to run slower or run before/after/concurrently with another
method in another thread---to {\em intervene} on discriminative predicates. For
example, to allow for return value alteration intervention, \asd modifies the
entire application by adding (1)~an optional parameter to each function, and
(2)~a conditional statement at the end of each function that specifies that
``if a value is passed to the optional parameter, the function should return
that specific value, and the actually computed value otherwise''. As another
example, the predicate ``there is a data race on X'' can be intervened by
delaying one access to X or by putting a lock around the code segments that
access X to prevent simultaneous access to X.
Figure~\ref{fig:runningExample}(d) shows how fault is injected by putting a
lock to intervene on the data race predicate of
Figure~\ref{fig:runningExample}(c).

\section{Real-world Concurrency Bug Characteristics} 
\looseness-1 Studies on real-world concurrency bug 
characteristics~\cite{DBLP:conf/asplos/LuPSZ08,
wong2009survey, DBLP:conf/icsm/VahabzadehF015} show that a vast majority of
root-causes can be captured with reasonably simple single predicates and hence
this assumption is very common in the SD
literature~\cite{statisticalDebuggingLiblit, Sober, crugLiblit}. Some notable
findings include: (1)~``97\% of the non-deadlock concurrency bugs are covered
by two simple patterns: \emph{atomicity violation} and \emph{order
violation}''~\cite{DBLP:conf/asplos/LuPSZ08}, (2)~``66\% of the non-deadlock
concurrency bugs involve only one variable''~\cite{DBLP:conf/asplos/LuPSZ08}
(3)~``The manifestation of 96\% of the concurrency bugs involves no more than
two threads.''~\cite{DBLP:conf/asplos/LuPSZ08}, (4)~``most fault localization
approaches assume that each buggy source file has exactly one line of faulty
code''~\cite{wong2009survey}, (5)~``The majority of flaky test bugs occur when
the test does not wait properly for asynchronous calls during the exercise
phase of testing.''~\cite{DBLP:conf/icsm/VahabzadehF015}, etc.

\section{Proof of Theorem~\ref{THEM2}}\label{theorem2Proof}
\begin{proof} After the first intervention, we get at least $\Big(\log{N
\choose D} - \log{N - S_1 \choose D} + 1 \Big)$ bits of information. Suppose
that there are $m$ interventions. Since after retrieving all information, the
remaining information should be $ \le 0$:
\allowdisplaybreaks
\begin{align*}
	&\log {N \choose D} - \sum_{i=1}^m\Big( \log{N{-}(i{-}1)S_1 \choose D} - \log{N{-}iS_1 \choose D} + 1\Big) \le 0\\
	&\implies \log{N{-}mS_1 \choose D} - m \le 0\\
	&\implies m \ge \log\frac{(N{-}mS_1)!}{D!(N{-}mS_1{-}D)!}\\
	&\implies m \ge \log\frac{(N{-}mS_1)^D}{D!} \;\;\;\;\; \text{[$\frac{(N{-}mS_1)!}{(N{-}mS_1{-}D)!} \approx (N - mS_1)^D$]}\\
	&\implies m \ge D\log(N{-}mS_1) - \log(D!)\\
	&\implies m \ge D\log N(1 - \frac{mS_1}{N}) - \log(D!)\\
	&\implies m \ge D\log N + D\log(1 - \frac{mS_1}{N}) - \log(D!)\\
	&\text{Since $\log$}\text{$(1 - x)\approx -x$ for small $x$; we assume $\frac{mS_1}{N}$ to be small:}\\
	&\implies m \ge D\log N - \frac{mDS_1}{N} - \log(D!)\\
	&\implies m \Big( 1 + \frac{DS_1}{N}\Big) \ge D\log N - \log(D!)\\
	&\implies m \Big( 1 + \frac{DS_1}{N}\Big) \ge \log \frac{N^D}{D!}\\
	&\implies m \Big( 1 + \frac{DS_1}{N}\Big) \ge \log \frac{N!}{D!(N-D)!}\;\;\;\;\; \text{[$N^D \approx \frac{N!}{(N{-}D)!}$]}\\
	&\implies m \ge \frac{\log{N \choose D}}{1 + \frac{DS_1}{N}}
\end{align*}
\end{proof}

\section{Proof of Theorem~\ref{THEM4}}\label{theorem4Proof}
\begin{proof}
\looseness-1 Since at least $S_2$ predicates are discarded during each causal
predicate discovery, and there are $D$ causal predicates, we compute the upper
bound of the number of required interventions:
\allowdisplaybreaks
\begin{align*}
	  &\sum_{i = 1}^D \log \Big(N - (i-1)S_2\Big)\\
	= &\sum_{i = 1}^D \log \Big(N\big(1 - \frac{(i-1)S_2}{N}\big)\Big)\\
	= &\sum_{i = 1}^D \log N + \sum_{i=1}^D\log \Big( 1 - \frac{(i-1)S_2}{N}\Big)\\
	\approx &\sum_{i = 1}^D \log N - \sum_{i=1}^D\frac{(i-1)S_2}{N} \text{\;\;[$\log (1 - x) \approx -x$ for small $x$]}\\
	= &D \log N - \frac{D(D-1)S_2}{2N}
\end{align*}
\end{proof}

\end{document}